\documentclass[letterpaper,11pt]{article}
\pdfoutput=1

\usepackage{jheppub}
\usepackage{multirow}

\usepackage{subfig}
\usepackage{xspace}
\usepackage[countmax]{subfloat}

\usepackage{amssymb}
\usepackage{amsmath}
\usepackage{cancel}
\usepackage{color}
\usepackage{braket}
\usepackage{graphicx}
\usepackage{multirow}
\usepackage{verbatim}
\usepackage{amsthm}
\usepackage{slashed}
\usepackage{wasysym}

\def\cA{\mathcal{A}}
\def\cB{\mathcal{B}}
\def\cC{\mathcal{C}} 
\def\cD{\mathcal{D}}

\def\cI{\mathcal{I}}

\def\cM{\mathcal{M}}

\def\cO{\mathcal{O}}

\def\cP{\mathcal{P}}

\newcommand{\Eq}[1]{Equation~\eqref{#1}}

\DeclareRobustCommand{\Sec}[1]{Sec.~\ref{#1}}
\DeclareRobustCommand{\Secs}[2]{Secs.~\ref{#1} and \ref{#2}}

\DeclareRobustCommand{\Tab}[1]{Table~\ref{#1}}

\DeclareRobustCommand{\Fig}[1]{Fig.~\ref{#1}}
\DeclareRobustCommand{\Figs}[2]{Figs.~\ref{#1} and \ref{#2}}
\DeclareRobustCommand{\Eq}[1]{Eq.~(\ref{#1})}

\def\be{\begin{equation}}
\def\ee{\end{equation}}

\newcommand{\ptveto}{p_{T}^{veto}}

\allowdisplaybreaks[3]

\newcommand{\eq}[1]{eq.~\eqref{eq:#1}}

\preprint{MIT--CTP 4548}

\title{Jet Vetoes Interfering with $H\rightarrow WW$ }

\author{Ian Moult}
\author{and Iain W.~Stewart}

\affiliation{Center for Theoretical Physics, Massachusetts Institute of Technology, Cambridge, MA 02139, USA}

\emailAdd{ianmoult@mit.edu}
\emailAdd{iains@mit.edu}

\abstract{Far off-shell Higgs production in $H \rightarrow WW,ZZ$, is a particularly powerful probe of Higgs properties, allowing one to disentangle Higgs width and coupling information unavailable in on-shell rate measurements. These measurements require an understanding of the cross section in the far off-shell region in the presence of realistic experimental cuts. We analytically study the effect of a $p_T$ jet veto on far off-shell cross sections, including signal-background interference, by utilizing hard functions in the soft collinear effective theory that are differential in the decay products of the $W/Z$.  Summing large logarithms of $M_{WW}/\ptveto$, we find that the jet veto induces a strong dependence on $M_{WW}$, modifying distributions in $M_{WW}$ and $M_T$. The example of $gg\rightarrow H \rightarrow WW$ is used to demonstrate these effects at next to leading log order. We also discuss the importance of jet vetoes and jet binning for the recent program to extract Higgs couplings and widths from far off-shell cross sections.}
\keywords{Higgs Physics, Hadronic Colliders, Jets}

\begin{document}

%{\flushright \\[-9ex]}
\maketitle

%%%%%%%%%%%%%%%%%%%%%%%%%%%%%%%%%%%%%%%%%%%%%%%%%%%%%%%%%%%%%%%%%%%%%%%%%%%%%%%%
\section{Introduction}
\label{sec:intro}
%%%%%%%%%%%%%%%%%%%%%%%%%%%%%%%%%%%%%%%%%%%%%%%%%%%%%%%%%%%%%%%%%%%%%%%%%%%%%%%%

With the recent discovery of a boson resembling a light Standard Model (SM) Higgs \cite{Aad:2012tfa,Chatrchyan:2012ufa,CMS:2014ega,Aad:2014aba,Khachatryan:2014ira}, a large program has begun to study in detail the properties of the observed particle \cite{Duhrssen:2004cv,Lafaye:2009vr,Bonnet:2011yx,Heinemeyer:2013tqa,Carmi:2012in,Carmi:2012yp,Djouadi:2012rh,Englert:2012wf,Klute:2012pu,Dobrescu:2012td,Plehn:2012iz,Corbett:2012ja,Belanger:2012gc,Farina:2012xp,Batell:2012ca,Espinosa:2012im,Espinosa:2012vu,Banerjee:2012xc,Barger:2012hv,Giardino:2012dp,Giardino:2012ww,Giardino:2013bma,Aad:2013wqa,Gainer:2013rxa,Belanger:2013xza,Belanger:2013kya,Ellis:2013lra,Cranmer:2013hia,Chen:2014pia}. Of fundamental interest are the couplings to SM particles and the total width of the observed boson, which is a sensitive probe of BSM physics  \cite{Djouadi:2005gj,Burgess:2000yq,Patt:2006fw,He:2011de,Raidal:2011xk,Englert:2011yb,Barbieri:2005ri}. Most studies have  focused on the extraction of Higgs properties from on-shell cross sections. In this case, the effect of jet vetoes and jet binning, which is required experimentally in many channels to reduce backgrounds, has been well studied theoretically \cite{Berger:2010xi,Banfi:2012yh,Banfi:2012jm,Becher:2012qa,Liu:2012sz,Tackmann:2012bt,Gangal:2013nxa,Becher:2013xia,Stewart:2013faa}. A jet veto, typically defined by requiring that there are no jets with $p_T \geq \ptveto$, introduces large logarithms, $\log(m_H/\ptveto)$, potentially invalidating the perturbative expansion, and requiring resummation for precise theoretical predictions. In this paper, we analytically study the effect of an exclusive jet $p_T$-veto on off-shell particle production, resumming logarithms of $\sqrt{\hat s}/\ptveto$, where $\sqrt{\hat s}$ is the invariant mass of the off-shell particle, or more precisely, $\sqrt{\hat s}$ is the invariant mass of the leptonic final state. We use $gg\rightarrow H \rightarrow WW$ as an example to demonstrate these effects, although the formalism applies similarly to $gg\rightarrow H\rightarrow ZZ$ if a jet veto is imposed. We find that the off-shell cross section is significantly suppressed by a jet veto, and that the suppression has a strong dependence on $\sqrt{\hat s}$. This results in a modification of differential distributions in $\sqrt{\hat s}$, or any transverse mass variable, in the case that the invariant mass cannot be fully reconstructed. The jet veto also has an interesting interplay with signal-background interference effects, which typically contribute over a large range of $\sqrt{\hat s}$. We use two cases, $m_H=126$ GeV, and $m_H=600$ GeV, to demonstrate the effect of the jet veto on the signal-background interference in $gg \rightarrow H \rightarrow WW$.

There exist multiple motivations why it is important to have a thorough understanding of the far off-shell region in Higgs production, and the impact of a jet $p_T$ veto on this region. As has been emphasized in a number of recent papers \cite{Dobrescu:2012td,Dixon:2013haa,Caola:2013yja,Campbell:2013una,Campbell:2013wga}, the separate extraction of the Higgs couplings and total width is not possible using only rate measurements for which the narrow width approximation (NWA) applies. In the NWA the cross section depends on the couplings and widths in the form
\be
\sigma^{\text{nwa}} \sim \frac{g_i^2 g_f^2}{\Gamma_H},
\ee
which is invariant under the rescaling
\be \label{eq:scaling_1}
 g_i \rightarrow \xi g_i,~~~~ \Gamma_H \rightarrow \xi^4 \Gamma_H,
\ee
  preventing their individual extraction from rate measurements alone.

The direct measurement of the width of the observed Higgs-like particle, expected to be close to its SM value of $\simeq 4$MeV, is difficult at the LHC, but is of fundamental interest as a window to new physics \cite{Djouadi:2005gj,Burgess:2000yq,Patt:2006fw,He:2011de,Raidal:2011xk,Englert:2011yb,Barbieri:2005ri}. It is also important for model independent measurements of the Higgs couplings. Proposals to measure the Higgs width include those that rely on assumptions on the nature of electroweak symmetry breaking \cite{Dobrescu:2012td}, direct searches for invisible Higgs decays \cite{CMS:2013bfa,CMS:2013yda,Bai:2011wz,Djouadi:2012zc,Englert:2012wf}, and a proposed measurement of the mass shift in $H\rightarrow \gamma \gamma$ relative to $H\rightarrow ZZ \rightarrow 4l$ caused by interference \cite{Dixon:2013haa}.

More recently, it has been proposed \cite{Caola:2013yja,Campbell:2013una,Campbell:2013wga} that the Higgs width can be bounded by considering the far off-shell production of the Higgs in decays to massive vector bosons. In this region there is a contribution from signal-background interference \cite{Campbell:2011cu,Kauer:2012ma,Passarino:2012ri,Kauer:2013qba}, and from far off-shell Higgs production \cite{Uhlemann:2008pm,Kauer:2012hd,Kauer:2013cga}. Far off-shell, the Higgs propagator is independent of $\Gamma_H$, giving rise to contributions to the total cross section that scale as
\be
\sigma^{\text{int}} \sim g_i g_f,~~~~ \sigma_H^{\text{off-shell}} \sim g_i^2 g_f^2,
\ee
for the signal-background interference and off-shell cross section respectively. The method proposed in \cite{Caola:2013yja} takes advantage of the fact that these components of the cross section scale differently than the NWA cross section. For example, in a scenario with large new physics contributions to the Higgs width, on-shell rate measurements at the LHC consistent with SM predictions enforce through \Eq{eq:scaling_1} that the Higgs couplings are also scaled as
$g_i\rightarrow g_i \left (    {\Gamma_H}/{\Gamma_H^{SM}} \right )^{1/4}$. The off-shell and interference contributions to the cross section are not invariant under this rescaling of the couplings, under which they are modified to
\begin{align}\label{Eq:Scaling}
 \sigma^{\text{int}} = \sqrt{\frac{\Gamma_H}{\Gamma^{SM}_H}}   \sigma_{SM}^{\text{int}},~~~~\sigma_H^{\text{off-shell}} =     \frac{\Gamma_H}{\Gamma^{SM}_H}      \sigma_{H, SM}^{\text{off-shell}}.
\end{align}
A measurement of the off-shell and interference cross section then allows for one to directly measure, or bound, the total Higgs width. 
This method is not completely model independent, indeed some of its limitations were recently discussed in \cite{Englert:2014aca}, along with a specific new physics model which decorrelated the on-shell and off-shell cross sections, evading the technique. However, interpreted correctly, this technique places restrictions on the Higgs width in many models of BSM physics. 
The study of the off-shell cross section as a means to bound the Higgs width was first discussed in the $H \rightarrow ZZ \rightarrow 4l$ channel \cite{Caola:2013yja,Campbell:2013una}, where the ability to fully reconstruct the invariant mass of the decay products allows for an easy separation of the on-shell and off-shell contributions. Recently, CMS has performed a measurement following this strategy and obtained a bound of $\Gamma_H \leq 4.2~ \Gamma_H^{SM}$ \cite{CMS:2014ala} .

The method was extended in \cite{Campbell:2013wga} to the $gg\rightarrow H\rightarrow WW \rightarrow l\nu l\nu$ channel. The WW channel has the advantage that the $2W$ threshold is closer than for $H\rightarrow ZZ$, as well as having a higher branching ratio to leptons, and a higher total cross section. It does however, also have the disadvantage of large backgrounds, which necessitate the use of jet vetoes, as well as final state neutrinos which prevent the reconstruction of the invariant mass. To get around the latter issue one can exploit the transverse mass variable
\be \label{eq:mt}
M_T^2=(E_T^{\text{miss}}+E_T^{\text{ll}})^2 -| \bf{p_T^{ll}+\bf{E}_T^{miss}}|^2,
\ee
which has a kinematic edge at $M_T=m_H$ for the signal. This variable was shown to be effective in separating the region where the off-shell and interference terms are sizeable, namely the high $M_T$ region, from the low $M_T$ region where on-shell production dominates, allowing for the extraction of a bound on the total Higgs width. Although the experimental uncertainties are currently large in the high $M_T$ region, the authors estimate that with a reduction in the background uncertainty to $\lesssim10\%$, the WW channel could be used to place a bound on the Higgs width competitive with, and complementary to the bound from the $H\rightarrow ZZ\rightarrow 4l$ channel. They therefore suggest a full experimental analysis focusing on the high-$M_T$ region of the $WW$channel. More generally, it was proposed in \cite{Coleppa:2014qja} that a similar method can also be used to probe couplings to heavy beyond the Standard Model states.

Independent of bounding the Higgs width, the study of the off-shell cross section opens up a new way to probe Higgs properties, which is particularly interesting as it probes particles coupling to the Higgs through loops over a large range of energies. Further benefits of the measurement of the off-shell cross section for constraining the parity properties of the Higgs, as well as for bounding higher dimensional operators were also discussed in \cite{Englert:2014aca,Gainer:2014hha}.

A full theoretical understanding of the far off-shell region, especially in the presence of realistic experimental cuts, is therefore well motivated to allow for a proper theoretical interpretation of the data, and of bounds on new physics. Indeed, the current limits on the Higgs width from the off-shell region are based on leading order calculations combined with a parton shower. There has recently been progress on the calculation of the perturbative amplitudes required for an NLO description of the off-shell cross section, including signal-background interference, with the calculation of the two loop master integrals with off-shell vector bosons \cite{Henn:2014lfa,Caola:2014lpa}. However, one aspect that has not yet been addressed theoretically is the effect of jet vetoes, and more generally jet binning, on far off-shell cross sections, and on the signal-background interference.

Jet vetoes and jet binning are used ubiquitously in LHC searches to reduce backgrounds. They are typically defined by constraining the $p_T$ of jets in the event. The $H\rightarrow WW$ channel is an example of such a search, where the exclusive zero jet bin, defined by enforcing that all jets in the event satisfy $p_T < \ptveto$, is used to reduce the large background from $t \bar t$ production. Indeed, the analysis of \cite{Campbell:2013wga} used the exclusive $N_{jet}=0$ bin in the large $M_T$ region to estimate the bound on the Higgs width achievable in the $H\rightarrow WW$ channel. Furthermore, the recent bound by CMS \cite{CMS:2014ala} of the Higgs width from the $H\rightarrow ZZ\rightarrow 2l2\nu$ channel used jet bins to optimize sensitivity, splitting data into exclusive $0$-jet and inclusive $1$-jet samples, which were each analyzed and then combined to give the limit. The proper interpretation of the off-shell cross section measurements requires understanding, preferably analytically, the impact of the jet veto and jet binning procedures.

As is well known, the jet veto introduces a low scale, typically $\ptveto\sim 25-30$GeV, into a problem which is otherwise characterized by the scale $Q$, of the hard collision. This causes large logarithms of the form $\alpha_s^n \log^m(Q/\ptveto),~m\leq 2n$, to appear in perturbation theory, forcing a reorganization of the perturbative expansion. Physically these logarithms arise due to constraints placed on the radiation in the event, which prevent a complete cancellation of real and virtual infrared contributions. A resummation to all orders in $\alpha_s$ is then required to make precise predictions. For the leading logarithms this resummation can be implemented by a parton shower. This approach is however difficult to systematically improve, and does not allow for higher order control of the logarithmic accuracy, or a systematic analysis of theoretical uncertainties in the correlations between jet bins. An alternative approach, which allows for the analytic resummation of large logarithms appearing in the cross section, is to match to the soft collinear effective theory (SCET) \cite{Bauer:2000ew,Bauer:2000yr,Bauer:2001ct,Bauer:2001yt,Bauer:2002nz}, which provides an effective field theory description of the soft and collinear limits of QCD. In SCET, large logarithms can be resummed through renormalization group evolution to desired accuracy, providing analytic control over the resummation. This framework also provides control over the theoretical uncertainties, including the proper treatment of correlations between jet bins~\cite{Berger:2010xi,Dittmaier:2012vm,Stewart:2011cf}.

The effect of jet vetoes on Higgs production in the on-shell region has attracted considerable theoretical interest \cite{Berger:2010xi,Banfi:2012yh,Banfi:2012jm,Becher:2012qa,Liu:2012sz,Tackmann:2012bt,Becher:2013xia,Stewart:2013faa}. For on-shell Higgs production, $Q\sim m_H$, and hence the resummation is of logarithms of the ratio $m_H/\ptveto$. The use of a jet clustering algorithm in the experimental analyses complicates resummation and factorization \cite{Ellis:2010rwa,Tackmann:2012bt}, and leads to logarithms of the jet radius parameter \cite{Tackmann:2012bt,Kelley:2012zs,Alioli:2013hba,Banfi:2012jm}. Current state of the art calculations achieve an NNLL$^\prime$+NNLO accuracy~\cite{Becher:2013xia,Stewart:2013faa}, along with the incorporation of the leading dependence on the jet radius, allowing for precise theoretical predictions in the presence of a jet veto. Such predictions are necessary for the reliable extractions of Higgs couplings from rate measurements. Indeed, the exclusive zero-jet Higgs cross section is found to decrease sharply as the $\ptveto$ scale is lowered.

In this paper we use SCET to analytically study the effect of a jet veto on off-shell cross sections. In particular, we are interested in processes with contributions from a large range of $\hat s$, where $\sqrt{\hat s}$ is the partonic centre of mass energy. In \Sec{sec:2}, we present a factorization theorem allowing for the resummation of large logarithms of the form $\log \sqrt{\hat s}/\ptveto$, in the cross section for the production of a non-hadronic final state. Working to NLL order, and using canonical scales, for simplicity, gives \cite{Banfi:2012yh}
\begin{align} \label{eq:intro_fact}
\frac{d\sigma_0^{\rm NLL}(\ptveto)}{d\hat s\, d\Phi}=  & \big|\cM_{ij}(\mu=\sqrt{\hat s},\Phi)\big|^2  \int\!\! dx_a dx_b f_i(x_a,\mu=\ptveto)f_j(x_b,\mu=\ptveto)    \\ 
& \times  \delta(x_a x_b E_\text{cm}^2-\hat s)   e^{-2{\rm Re} K^i_{\rm NLL}(\sqrt{\hat s},\, \ptveto)} \,,
 \nonumber
\end{align}
where $\sigma_0(\ptveto)$ is the exclusive zero-jet cross section. In this formula, $f_i$, $f_j$ are the parton distribution functions (PDFs) for species $i,j$, $\cM_{ij}$ is the hard matrix element, $\Phi$ is the leptonic phase space, and $E_{\text{cm}}$ is the hadronic centre of mass energy. $K^i_{\text{NLL}}$ is a Sudakov factor, defined explicitly in \Sec{sec:2}, which depends only on the identity of the incoming partons. The form of \Eq{eq:intro_fact} shows that the effect of the jet veto can be captured independent of the hard underlying process, which enters into \Eq{eq:intro_fact} only through $\cM$. At higher logarithmic order a dependence on the jet algorithm is also introduced, but the ability to separate the effect of the jet veto from the particular hard matrix element using the techniques of factorization remains true, and allows one to make general statements about the effect of the jet veto.

The resummation of the large logarithms, $\log \sqrt{\hat s}/\ptveto$, introduced by the jet veto leads to a suppression of the exclusive zero-jet cross section, evident in \Eq{eq:intro_fact} through the Sudakov factor, and familiar from the case of on-shell production. The interesting feature in the case of off-shell effects is that this suppression depends on $\sqrt{\hat s}$. For example, when considering off-shell Higgs production, or signal-background interference, which contribute over a large range of $\sqrt{\hat s}$, the jet veto re-weights contributions from different $\sqrt{\hat s}$ regions in a strongly $\sqrt{\hat s}$ dependent manner. In particular, this modifies differential distributions in $\sqrt{\hat s}$, or any similar variable, such as $M_T$. This is of particular interest for the program to place bounds on the Higgs width using the off-shell cross section in channels which require a jet veto, as this procedure requires an accurate description of the shape of the differential cross section. Furthermore, the jet veto has an interesting effect on the signal-background interference, which often exhibits cancellations from regions widely separated in $\sqrt{\hat s}$. The study of these effects is the subject of this paper.

Our outline is as follows. In \Sec{sec:2} we review the factorization theorem for the exclusive zero jet bin, with a jet veto on the $p_T$ of anti-$k_T$ jets, focussing on the dependence on $\sqrt{\hat s}$. \Sec{general} describes the generic effects of jet vetoes on off-shell production, including the dependence on the jet veto scale, the identity of the initial state partons and the hadronic centre of mass energy. In particular, we show that off-shell production in the exclusive zero-jet bin is suppressed by a strongly $\sqrt{\hat s}$ dependent Sudakov factor, and comment on the corresponding enhancement of the inclusive 1-jet cross section. In \Sec{sec:WW} we perform a case study for the $gg\rightarrow H\rightarrow WW\rightarrow l\nu l\nu$ process, resumming to NLL accuracy the off-shell cross section including the signal-background interference. For the signal-background interference, we consider two Higgs masses, $m_H=125$ GeV and $m_H=600$ GeV, whose interference depends differently on $\sqrt{\hat s}$, to demonstrate different possible effects of the jet veto on the signal-background interference. Since $\sqrt{\hat s}$ is not experimentally reconstructible for $H\to WW$, in \Sec{sec:mt_126}, we demonstrate the suppression as a function of $M_T$ caused by the jet-veto restriction. In \Sec{sec:bound} we discuss the effect of the jet veto and jet binning on the extraction of the Higgs width from the off-shell cross section in $H\rightarrow WW$ (commenting also on $H\to ZZ$). We conclude in \Sec{sec:conclusion}.

\section{Cross Sections with a Jet Veto: A Review}\label{sec:2}

In this section we review the factorization theorem, in the SCET formalism, for $pp \rightarrow L+0$-jets, where L is a non-hadronic final state. We consider a jet veto defined by clustering an event using an anti-$k_T$ algorithm with jet radius $R$ to define jets, $J$, and imposing the constraint that $p_T^J <\ptveto$ for all jets in the event. This is the definition of the jet veto currently used in experimental analyses, with the experimental value of $\ptveto$ typically $25-30$ GeV, and $R\simeq 0.4\text{-}0.5$. 

\subsection{Factorization Theorem}
Following the notation of \cite{Stewart:2013faa}, the factorization theorem for $pp \rightarrow L+0$-jets with a jet veto on $p_T$ can be computed in the framework of SCET. For a hard process where $L$ has invariant mass $\sqrt{\hat s}$ (on-shell or off-shell), we have 
\begin{align} \label{eq:fact}
&\frac{d\sigma_0(\ptveto)}{d \hat s}=\int d\Phi dx_a dx_b ~\delta(x_a x_bE_{cm}^2-\hat s) \sum_{i,j} H_{ij}(\sqrt{\hat s},\Phi,\mu) B_i(\sqrt{\hat s},\ptveto,R,x_a,\mu,\nu) \nonumber \\
&\times  B_j(\sqrt{\hat s},\ptveto,R,x_b,\mu,\nu)  S_{ij}(\ptveto,R,\mu,\nu)  
\nonumber \\
&
+\frac{d\sigma_0^{Rsub}(\ptveto,R)}{d \hat s}+\frac{d\sigma_0^{ns}(\ptveto,R,\mu_{ns})}{d\hat s}.
\end{align}
In this formula, $\Phi$ denotes the leptonic phase space, $i,j$ denote the initial partonic species, $H_{ij}$ is the hard function for a given partonic channel, $B_i$ are the beam functions which contain the PDFs, and $S_{ij}$ is the soft function, each of which will be reviewed shortly. Since this factorization theorem applies to the production of a color singlet final state, we either have $i=j=g$, or $i=q$, $j=\bar q$.   \Eq{eq:fact} is written as the sum of three terms. The first term in \Eq{eq:fact} contains the singular logarithmic terms, which dominate as $\ptveto \rightarrow 0$, or in the case of off-shell production that we are considering, as $\hat s \rightarrow \infty$, with $\ptveto$ fixed. The second term, $\sigma_0^{Rsub}$, contains corrections that are polynomial in the jet radius parameter $R$, and $\sigma_0^{ns}$ contains non-singular terms which vanish as $\ptveto \rightarrow 0$, and are suppressed relative to the singular terms when the ratio $\ptveto/\sqrt{\hat s}$ is small. 

The factorization theorem allows for each component of \Eq{eq:fact} to be calculated at its natural scale, and evolved via renormalization group evolution (RGE) to a common scale, resumming the large logarithms of $\ptveto/\sqrt{\hat s}$. For the case of a veto on the jet $p_T$, the factorization follows from SCET$_{\text{II}}$, where the RGE is in both the virtuality scale, $\mu$, and rapidity scale, $\nu$ \cite{Chiu:2011qc,Chiu:2012ir}. In this section, we will briefly summarize the components of the factorization theorem with a particular focus on their dependence on the underlying hard matrix element, the identity of the incoming partons, the jet algorithm, and the jet veto measurement. We will also review their RGE properties. Further details, including analytic expressions for the anomalous dimensions, can be found in \cite{Tackmann:2012bt,Stewart:2013faa,Berger:2010xi}, and references therein.

\subsection*{Soft Function}

The soft function $S_{ij}(\ptveto,R,\mu,\nu)$ describes the soft radiation from the incoming partons $i,j$ which are either both gluons or both quarks. It is defined as a matrix element of soft Wilson lines along the beam directions, with a measurement operator, $\cM^{jet}$, which enforces the jet veto condition:
\be
S_{ii}(\ptveto, R, \mu, \nu) = \bra{0} Y_{n_b} Y^\dagger_{n_a} \cM^{jet}(\ptveto,R)Y_{n_a} Y^\dagger_{n_b} \ket{0}.
\ee
The soft function depends only on the identity of the incoming partons, through the representation of the Wilson lines, which has not been made explicit.  It also depends on the definition of the jet veto through the measurement function.

The soft function is naturally evaluated at the soft scale $\mu_S \sim \ptveto$, and $\nu_S \sim \ptveto$, and satisfies a multiplicative renormalization in both $\mu$ and $\nu$. The solution is given by
\begin{align}
&S_{ii}(\ptveto,R,\mu,\nu)=S_{ii}(\ptveto,R,\mu_S,\nu_S) \exp \left[\log\frac{\nu}{\nu_S}\gamma_\nu^i(\ptveto,R,\mu_S)  \right] 
\exp \bigg[ \int\limits_{\mu_S}^{\mu} \frac{d\mu'}{\mu'} \gamma^i_S(\mu',\nu) \bigg] \,.
\end{align}
Further details including expressions for the anomalous dimensions are given in \cite{Stewart:2013faa}.

In the case of interest, where the jets are defined using a clustering algorithm with a finite R, the soft function also contains clustering logarithms from the clustering of correlated soft emissions, which first arise at NNLL. These appear in the cross section as logarithms of the jet radius, $\log(R)$, but are not resummed by the RGE. For experimentally used values of R, the first of these logarithms is large \cite{Banfi:2012jm}, while the leading $\cO(\alpha_s^3)$ term  was recently calculated and found to be small \cite{Alioli:2013hba}. We therefore treat these $\log(R)$ factors in fixed order perturbation theory. We discuss the impact of these logarithms on our results in \Sec{sec:conv}.

\subsection*{Beam Function}
The beam function \cite{Fleming:2006cd,Stewart:2009yx,Stewart:2010qs}, $B_i$, describes the collinear initial state-radiation from an incoming parton, $i$, as well as its extraction from the colliding protons through a parton distribution function. The beam function depends only on the identity of the incoming parton $i$, and the measurement function. 

In the case of a $\ptveto$, the beam function can be calculated perturbatively by matching onto the standard PDFs at the beam scale $\mu_B\sim \ptveto$, $\nu_B\sim \sqrt{\hat s}$:
\begin{align}
B_i(\sqrt{\hat s},\ptveto,R,x,\mu_B,\nu_B)=\sum_j \int\limits_x^1 \frac{dz}{z} 
  \cI_{ij}\big(\sqrt{\hat s},\ptveto,R,z,\mu_B,\nu_B\big) f_j\Big(\frac{x}{z},\mu_B\Big)
\end{align}
The lowest order matching coefficient is
\be
\cI_{ij}=\delta_{ij} \delta(1-z)
\ee
so that to leading order the beam function is simply the corresponding PDF, but evaluated at the beam scale $\mu_B\simeq \ptveto$. This was seen explicitly in the NLL expansion of \Eq{eq:intro_fact}. Higher order matching coefficients involve splitting functions, allowing for a mixing between quarks and gluons. This matching procedure corresponds to the measurement of the proton at the scale $\mu_B \sim \ptveto$ by the jet veto. Above the scale $\mu_B$, the beam function satisfies a multiplicative RGE in both virtuality, $\mu$ and rapidity, $\nu$, describing the evolution of an incoming jet for the off-shell parton of species $i$. Unlike the RGE for the PDFs, the RGE for the beam function leaves the identity and momentum fraction of the parton unchanged. The solution to the RGE is given by
\begin{align}
&B_{i}(\sqrt{\hat s},\ptveto,R,x,\mu,\nu)=B_{i}(\sqrt{\hat s},\ptveto,R,x,\mu_B,\nu_B) \exp \left[\frac{1}{2}\log\frac{\nu_B}{\nu}\gamma_\nu^i(\ptveto,R,\mu_B)  \right] \nonumber \\
&\times \exp \bigg[ ~\int\limits_{\mu_B}^{\mu} \frac{d\mu'}{\mu'} \gamma^i_B(\sqrt{\hat s},\mu',\nu) ~\bigg],
\end{align}
which resums the logarithmic series associated with the collinear radiation from the incoming parton. Further details and expressions for the anomalous dimensions are again given in \cite{Stewart:2013faa}.

As with the soft function, the beam function also contains logarithms and polynomial dependence  on the jet radius, $R$, from the clustering of collinear emissions. These logarithms can be numerically significant, but are not resummed by the RGE. We again treat these terms in fixed order perturbation theory.

\subsection*{Hard Function}
The hard function $H_{ij}$ encodes the dependence of the singular term of \Eq{eq:fact} on the underlying hard partonic matrix element of the $pp \rightarrow L+0$-jets process. It can be obtained by matching QCD onto an appropriate SCET operator at the scale $\sqrt{\hat s}$, giving a Wilson coefficient, $C_{ij}$. The Wilson coefficient satisfies a standard RGE in virtuality, allowing it to be evolved to the scale $\mu$. The hard function is then given by the square of the Wilson coefficient
\be
H_{ij}(Q,\mu)=|C_{ij}(Q,\mu)|^2 \,,
\ee
where $Q$ denotes dependence on all variables associated with the final leptons as well as parameters like the top-mass, and the Higgs and $W/Z$ masses and widths. The solution to the RGE equation for the hard function is
\begin{align}
  H_{ij}(Q,\mu) = H_{ij}(Q,\mu_H)  \Big| e^{-K^i(\sqrt{\hat s},\mu_H,\mu)} \Big|^2 
\,,
\end{align}
where the Sudakov form factor is
\begin{align} \label{eq:K}
  K^i(\sqrt{\hat s},\mu_H,\mu) &=\int_{\mu_H}^{\mu} \frac{d\mu'}{\mu'}  \gamma_H^i(\sqrt{\hat s},\mu') 
  \nonumber \\
  &= 2 K_{\Gamma_{\rm cusp}^i}(\mu_H,\mu) - K_{\gamma_H^i}(\mu_H,\mu) 
    - \ln\Big(\frac{-\hat s-i0}{\mu_H^2}\Big) \:  \eta_{\Gamma_{\rm cusp}^i}(\mu_H,\mu) \,.
\end{align}
Here the integrals involve the $\beta$-function and anomalous dimensions
\begin{align}  \label{eq:Ks}
K_{\Gamma^i_{\rm cusp}} &= \int_{\alpha_s(\mu_H)}^{\alpha_s(\mu)}  \frac{d\alpha_s}{\beta(\alpha_s)} \Gamma^i_{\rm cusp}(\alpha_s) \int_{\alpha_s(\mu_H)}^{\alpha_s} \frac{d\alpha_s'}{\beta(\alpha_s')} 
  \,,  
& \eta_{\Gamma_{\rm cusp}^i} &= \int_{\alpha_s(\mu_H)}^{\alpha_s(\mu)}  \frac{d\alpha_s}{\beta(\alpha_s)} \Gamma_{\rm cusp}^i(\alpha_s) \,,
\nonumber \\
  K_{\gamma_H^i} &= \int_{\alpha_s(\mu_H)}^{\alpha_s(\mu)}  \frac{d\alpha_s}{\beta(\alpha_s)} \gamma_H^i(\alpha_s)
 \,,
\end{align}
where  the channel $i$ is either for quarks or gluons. Here the cusp and regular anomalous dimensions are  $\Gamma_{\rm cusp}^i(\alpha_s) = \sum_{k=0}^\infty \Gamma^i_k (\alpha_s/4\pi)^{k+1}$,
$\gamma_H^i(\alpha_s) = \sum_{k=0}^\infty \gamma^i_k (\alpha_s/4\pi)^{k+1}$, respectively, and the $\beta$-function is $\beta(\alpha_s) = -2 \alpha_s \sum_{k=0}^\infty \beta_k (\alpha_s/4\pi)^{k+1}$ so $\beta_0=11 C_A/3-2n_f/3$.

Explicit results for the functions in Eq.~(\ref{eq:Ks}) can be found for example in Ref.~\cite{Berger:2010xi}.
Since we will be considering far off-shell production, and including signal-background interference effects, which have not been discussed in SCET factorization theorems before, we will discuss in more detail the definition of the hard function for the specific case of $gg\rightarrow l\nu l\nu$ in \Sec{sec:hardfunc}.

The beam and soft functions are universal, depending only on the given measurement and the identity of the incoming partons, it is the hard function that needs to be calculated separately for different processes. The beam and soft functions are known to NNLL for the case of a jet veto defined using a cut on $p_T$, and it is the hard coefficient that prevents resummation to NNLL for several cases of interest. In particular, since we are interested here in the case of off-shell production, one needs the full top mass dependence of loops, significantly complicating the computation. Indeed, for the case of signal-background interference for $gg \rightarrow H\rightarrow WW\rightarrow l\nu l\nu$, only the leading order hard function is known \cite{Campbell:2011cu}, while for direct gluon-fusion Higgs production, analytic results exist for the NLO virtual corrections including quark mass dependence \cite{Harlander:2005rq}. This restricts our predictions to NLL accuracy for signal-background interference for $gg \rightarrow H\rightarrow WW\rightarrow l\nu l\nu$.

\subsection*{Non-Singular Terms}
The non-singular term $\sigma_0^{ns}(\ptveto,R,\mu_{ns})$ is an additive correction to the factorization theorem, containing terms that vanish as $\ptveto \rightarrow 0$. This term scales as $\ptveto/\sqrt{\hat s}$. The non-singular piece is important when $\ptveto$ is of the same order as $\sqrt{\hat s}$, where both singular and non-singular pieces contribute significantly to the cross section.  In this paper, we will be focusing on the effect of a jet veto on far off-shell effects, and we will therefore always be considering the case that $\ptveto \ll \sqrt{\hat s}$. We will therefore not discuss the non-singular pieces of the cross section, and focus on the singular contributions.

\subsection*{Uncorrelated Emissions}
Beginning with two emissions, the jet algorithm can cluster uncorrelated emissions from the soft and collinear sectors \cite{Tackmann:2012bt,Becher:2012qa,Banfi:2012jm}. This produces terms proportional to powers of $R^2$, which can formally be treated as power corrections for $R\ll 1$, and are included in $\sigma_0^{Rsub}$. For the jet radii of $0.4 \text{-}0.5$ used by the experimental collaborations, these effects are numerically very small, especially compared to the $\log R$ terms from correlated emissions. We make use of the expressions from \cite{Stewart:2013faa}.

\subsection{Expansion to NLL}\label{sec:NLL}

It is useful to consider the factorization theorem at NLL order with canonical scale choices, to see the main factors that control its behaviour. The result at NLL was first given in  \cite{Banfi:2012yh} for on-shell production with $\sqrt{\hat s}=m_H$. Allowing for off-shell production, and using canonical scales, the cross section with a $\ptveto$ cut is given at NLL by
\begin{align} \label{eq:eqNLL}
\frac{d\sigma_0^{\rm NLL}(\ptveto)}{d\hat s\, d\Phi}=  & \big|\cM_{ij}(\mu=\sqrt{\hat s},\Phi)\big|^2  \int\!\! dx_a dx_b f_i(x_a,\mu=\ptveto)f_j(x_b,\mu=\ptveto)    \\ 
& \times  \delta(x_a x_b E_\text{cm}^2-\hat s)   e^{-2{\rm Re} K^i_{\rm NLL}(\sqrt{\hat s},\, \ptveto)} \,,
 \nonumber
\end{align}
where $\Phi$ are phase space variables for the final state leptonic decay products.
In this equation, $f_i$ and $f_j$ are the appropriate PDFs, for example, they are both $f_g$ for the case of gluon-fusion since direct contributions from the quark PDFs do not enter until NNLL order. For a partonic center of mass energy $\sqrt{\hat s}$, \Eq{eq:eqNLL} resums to NLL accuracy the logarithms of $ \sqrt{\hat s}/ \ptveto$. \Eq{eq:eqNLL} does not include the non-singular contribution to the cross section. As discussed previously, in the far off-shell region, $\ptveto \ll \sqrt{\hat s}$, and the singular contributions to the cross section dominate. It should also be emphasized that at NLL one is not sensitive to the jet algorithm or jet radius, as at $\cO(\alpha_s)$ there is only a single soft or collinear emission. Although the R dependence is important for accurate numerical predictions, it does not effect the qualitative behaviour of the jet veto. The R dependence appears in the factorization theorem at NNLL.

The only dependence on the hard partonic process in \Eq{eq:eqNLL} is in the matrix element $\cM_{ij}(\hat s)$. The Sudakov form factor $K^i$ given in Eq.~(\ref{eq:K}) arises from restrictions on real radiation in QCD, and depends only on the identity of the incoming partons. At NLL the Sudakov factor is given by
\begin{align}  \label{KNLL}
K_{\rm NLL}^i(\sqrt{\hat s},\mu_H,\mu) &= - \frac{\Gamma_0^i}{2\beta_0^2} \bigg\{ 
   \frac{4\pi}{\alpha_s(\mu_H)} \Big( 1-\frac{1}{r} - \ln r\Big) + \bigg( \frac{\Gamma_1^i}{\Gamma_0^i}-\frac{\beta_1}{\beta_0}\bigg) (1-r+\ln r) + \frac{\beta_1}{2\beta_0} \ln^2r \bigg\} 
\nonumber \\
  & \quad + \frac{\gamma_0^i}{2\beta_0}\ln r
   + \ln\Big(\frac{-\hat s-i0}{\mu_H^2}\Big) \frac{\Gamma_0^i}{2\beta_0} \bigg\{ \ln r +
  \frac{\alpha_s(\mu_H)}{4\pi} \bigg( \frac{\Gamma_1^i}{\Gamma_0^i}-\frac{\beta_1}{\beta_0}\bigg)(r-1)\bigg\}
  \,,
\end{align}
where $r=\alpha_s(\mu)/\alpha_s(\mu_H)$. The form in Eq.~(\ref{KNLL}) allows for the use of complex scales, such as $\mu_H=-i \sqrt{\hat s}$, to minimize the appearance of large $\pi^2$ factors in the Hard function.  On the other hand, with canonical scales we would take $K_{\rm NLL}^i=K_{\rm NLL}^i(\sqrt{\hat s},\ptveto)$.  At LL order the terms with $\Gamma_1$, $\beta_1$, and $\gamma_0$ do not yet contribute and using the LL running coupling we can write ${\rm Re} K_{\rm LL}^i(\sqrt{\hat s},\ptveto)=-(4 C/\beta_0) \ln\sqrt{\hat s}/\ptveto\, [
1 + \ln(1-2\lambda)/(2\lambda)]$ where $\lambda= \alpha_s(\sqrt{\hat s})\frac{\beta_0}{4\pi} \ln\sqrt{\hat s}/\ptveto$.
For gluon-fusion, $C=C_A=3$, whereas for a quark-antiquark initial state, $C=C_F=4/3$.

There are two important features of the expression in Eq.~(\ref{eq:eqNLL}) compared with the case of no jet veto. First, the PDFs are evaluated at the scale $\mu=\ptveto$ instead of $\mu=\sqrt{\hat s}$. Secondly, the cross section is multiplied by a Sudakov factor, which depends on logs of the ratio $\sqrt{\hat s}/\ptveto$. These have a strong impact on the cross section, which will be the focus of \Sec{general}.

\section{Jet Vetoes and Off-Shell Effects}\label{general}

In this section we will discuss quite generally the effect of jet vetoes on off-shell cross sections.  We focus on the dependence on the identity of the initial state partons, and the relation between the exclusive $0$-jet and inclusive $1$-jet bins. We conclude with a discussion of the dependence on the hadronic centre of mass energy. For simplicity, in this section we will use the NLL expansion of \Eq{eq:eqNLL} with canonical scale choices. The NLL expansion demonstrates the essential features that persist at higher logarithmic order, and makes transparent how these effects depend on various parameters of interest. This serves for the purpose of demonstrating the generic effects of jet vetoes, and their dependencies. In \Sec{sec:WW}, we will perform a more detailed study for the specific case of $gg\rightarrow H\rightarrow WW$. 

Unlike on-shell effects, which contribute to the cross section over a small region in $\sqrt{\hat s}$, of order the width, off-shell effects, including signal-background interference and off-shell production, typically contribute over a large range of values of $\sqrt{\hat s}$. In this case the $\sqrt{\hat s}$ dependence of the jet veto suppression can produce interesting effects. In particular, it modifies differential distribution in $\sqrt{\hat s}$, or any substitute such as $M_T$ in cases where the full invariant mass cannot be reconstructed, such as $H\rightarrow WW$. Furthermore, for signal-background interference, the $\sqrt{\hat s}$ dependence of the jet veto suppression can cause an enhancement or suppression of the interference relative to the on-shell contribution to the cross section, or enhance/suppress interference contributions with different signs relative to one another. 

With this motivation, we now study the $\sqrt{\hat s}$ dependence of the jet veto suppression to the exclusive zero jet cross section using the NLL expression of \Sec{sec:NLL}. The benefit of the factorized expression is that this discussion can be carried out essentially independent of the matrix element prefactor $|{\cal M}_{ij}|^2$. From \Eq{eq:eqNLL}, the NLL cross section is modified compared with the LO cross section, only by the evaluation of the PDFs at the jet veto scale, and by the Sudakov factor, which is a function of $\sqrt{\hat s}$. To study the suppression due to the jet veto as a function of $\sqrt{\hat s}$, we will therefore consider
\begin{align} \label{eq:ratio}
E_0(\hat s)=\left .   \left (  \frac{  d\sigma_0^{\rm NLL}(\ptveto)  } {d\sqrt{\hat s}}   \right )   \middle /      \left (  \frac{d\sigma}{d\sqrt{\hat s}} \right)      \right . ,
\end{align}
where $\sigma_0^{NLL}(\ptveto)$ is the NLL exclusive zero jet cross section. In \Eq{eq:ratio}, the cross section in the denominator is evaluated to LO, namely to the same order as the matrix element that appears in \Eq{eq:eqNLL} for the NLL resummed cross section. When forming this combination, one could choose to evaluate the denominator at various orders, for example using the full NLO result calculated without the $\ptveto$. Since NLO corrections are typically large, especially for gluon initiated processes, this would typically decrease the above ratio. However, we have in mind an application to processes, such as signal-background interference in $gg\rightarrow l\nu l\nu$, for which the NLO corrections are not yet known, so that current calculations are restricted to LO results. In this case, we can incorporate the effect of the jet veto at NLL using \Eq{eq:eqNLL}, and the ratio of \Eq{eq:ratio} will characterize the effect of the resummation compared to previous calculations in the literature \cite{Campbell:2011cu,Campbell:2013una,Campbell:2013wga}. This approach also has the benefit that it can be done independent of the particular matrix element, as the NLO corrections are clearly process dependent. However, all of the general features described in this section persist to NNLL resummation, as will be demonstrated in \Sec{sec:WW}. As was mentioned previously, at NLL one doesn't have sensitivity to the jet radius R. While this dependence is important for precise predictions, it does not dominate the behaviour of the jet vetoed cross section as a function of $\hat s$, or modify in any way the conclusions of this section. 

For numerical calculations in this section we use the NLO PDF fit of Martin, Stirling, Thorne and Watt \cite{Martin:2009iq} with $\alpha_s(m_Z)=0.12018$. Unless otherwise stated, we use a hadronic center of mass energy of $E_\text{cm}=8$ TeV. In \Sec{sec:Ecm} we discuss the dependence on the $E_{\text{cm}}$, comparing behaviour at $8,13,$ and $100$ TeV.

\begin{figure}
\begin{center}
\subfloat[]{\label{fig:glumi_ratio}
\includegraphics[width=7.0cm]{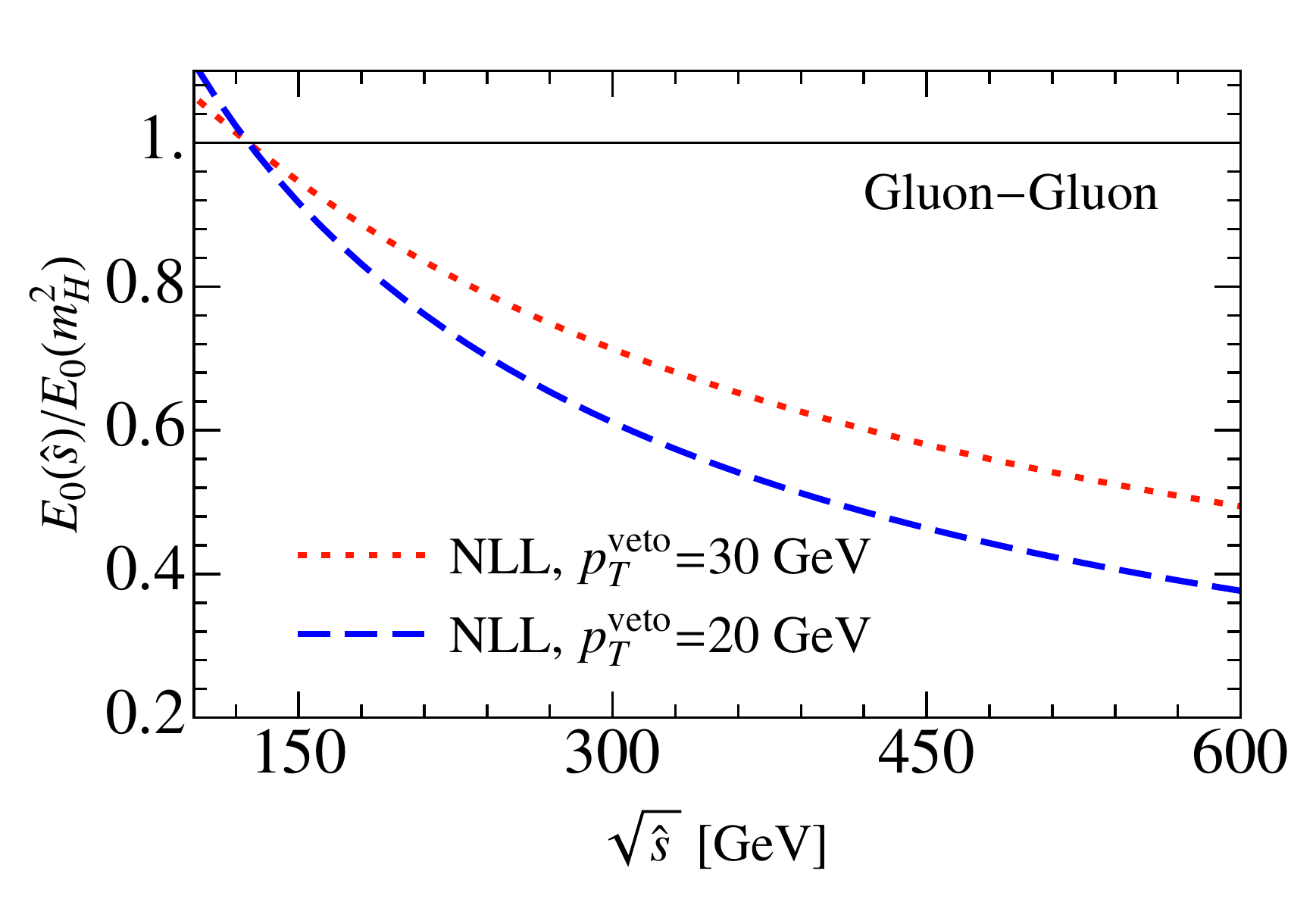}
}
$\qquad$
\subfloat[]{\label{fig:qlumi_ratio}
\includegraphics[width=7cm]{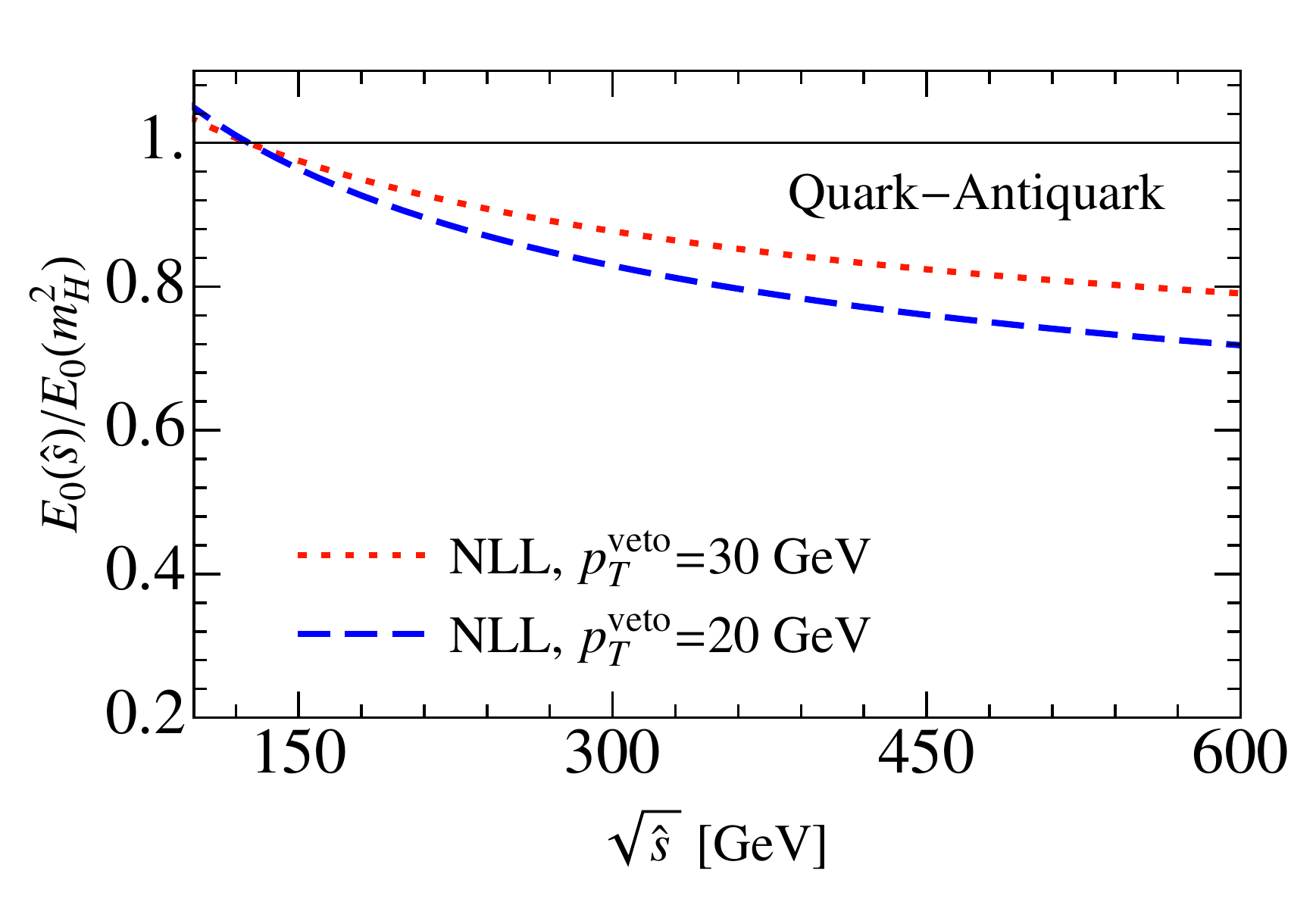}
}
\end{center}
\caption{The ratio  $E_0(\hat s)/E_0(m^2_H)$, for both a gluon-gluon initiated process in (a), and a quark-antiquark initiated process in (b). In both cases we consider $\ptveto=20,30$ GeV. The jet veto causes an $\hat s$ dependent suppression, which is significantly stronger for initial state gluons than initial state quarks, due to the larger colour factor appearing in the Sudakov.}
\label{fig:lumi_ratio}
\end{figure}

In \Fig{fig:glumi_ratio} we demonstrate the effect of the jet veto for a gluon-gluon initial state, as a function of $\sqrt{\hat s}$.\footnote{Note that a similar effect was considered in \cite{Berger:2010xi} which performed resummation for gluon fusion Higgs production with a veto on the global beam thrust event shape, as a function of the Higgs mass.} We plot the ratio $E_0( \hat s)/E_0(m^2_H)$, for $m_H=126$ GeV. We have chosen to plot this particular ratio to focus on the $\hat s$ dependence, rather than the impact that the jet veto has on the on-shell Higgs production cross section which is given by $E_0(m^2_H)$. The ratio $E_0( \hat s)/E_0(m^2_H)$ describes the impact of the jet-veto for  off-shell effects relative to its impact for on-shell production.   It will also be useful when discussing the impact on Higgs width bounds in \Sec{sec:bound}. \Fig{fig:glumi_ratio} shows that the suppression of the exclusive zero-jet cross section has a strong dependence on $\hat s$. The comparison between $\ptveto=20$ GeV, and $\ptveto=30$ GeV shows that a lower cut on the $p_T$ of emissions causes a more rapid suppression, as expected. We have chosen to use the values $\ptveto=20,~30$ GeV, because CMS currently uses $\ptveto=30$ GeV, and although the ATLAS collaboration uses $\ptveto=25$ GeV, the $\ptveto=20$ GeV cut demonstrates the effects of a fairly extreme jet veto. \Fig{fig:glumi_ratio} demonstrates that at scales of $\sqrt{\hat s} \simeq 500$ GeV, the suppression relative to that for on-shell production is of order $50\%$.

\subsection{Quarks vs. Gluons}\label{sec:qvsg}
We now consider the difference in the jet veto suppression for quark initiated and gluon initiated processes. This is relevant in the case where multiple partonic channels contribute to a given process, or if the signal and background processes are predominantly from different partonic channels. This is the case for both $gg\rightarrow H\rightarrow WW,~ZZ$, which have large $q\bar q$ initiated backgrounds. The factorization theorem in \Eq{eq:fact} allows one to easily study the dependence of the jet veto suppression on the identity of the incoming partons, which is carried by the hard, beam, and soft functions. The difference in the suppression arises from the differences in the anomalous dimensions, where for the 0-jet cross section, they involve $C_F$ for quarks, and $C_A$ for gluons. The clustering and correlation logarithms are also multiplied by the colour factors $C_F$ and $C_A$. This phenomenon is similar to quark vs. gluon discrimination for jets~\cite{Altheimer:2013yza}, where the same factors of $C_F$ and $C_A$ appear in the Sudakov and allow one to discriminate between quark and gluon jets. However, in this case, the discrimination is between incoming quarks and gluons. 

Comparing \Fig{fig:glumi_ratio} and \Fig{fig:qlumi_ratio}, we see a significant difference between a gluon-gluon and quark-antiquark initial state. The jet veto suppression increases more rapidly with $\hat s$ in the case of gluon-fusion induced processes than quark anti-quark induced processes. The suppression due to the jet veto being approximately twice as large for the case of gluon-fusion as for quark-antiquark fusion, for the values considered in \Fig{fig:lumi_ratio}. (Note that for the quark-antiquark initial state, we have used the up quark for concreteness, however, the result is approximately independent of flavour for the light quarks, with the suppression being dominated by the flavour independent Sudakov factor. A small dependence on flavour comes from the scale change in the PDF.) The effect of the jet veto is therefore of particular interest for gluon initiated processes, such as Higgs production through gluon-gluon fusion, to be discussed in \Sec{sec:WW}. This difference in the suppression is interesting for a proper analysis of the backgrounds for $H\rightarrow WW,~ZZ$ in the off-shell region, and deserves further study since one may wish to vary $\ptveto$ as a function of $\sqrt{\hat s}$ or $M_T$.

\subsection{Inclusive 1-Jet Cross Section}\label{sec:1jet}
We have up to this point focused on the exclusive zero jet cross section. However, since the total inclusive cross section is unaffected by the jet veto, the inclusive 1-jet cross section has the same logarithmic structure as the exclusive zero-jet cross section, and can be related to the exclusive zero jet cross section by 
\begin{equation}
\frac{d\sigma_{\geq 1}(\ptveto)}{d \sqrt{\hat s}}=\frac{d\sigma}{d \sqrt{\hat s}}-\frac{d\sigma_{0}(\ptveto)}{d \sqrt{\hat s}}.   \label{eq:1jet}
\end{equation}
In this equation, $\sigma_{\geq 1}(\ptveto)$ is the inclusive 1-jet cross section defined by requiring at least one jet with $p_T\geq \ptveto$, $\sigma_0$ is the exclusive zero-jet cross section and $\sigma$ is the inclusive cross section. This relation allows us to discuss the properties of the inclusive 1-jet bin as a function of $\hat s$ using the factorization theorem for the exclusive 0-jet cross section. Of particular interest is the split of the total cross section between the exclusive zero-jet bin and the inclusive 1-jet bin, and the migration between the two bins as a function of $\hat s$. This relation also implies a correlation between the theory uncertainties for the resummation for the two jet bins, which is important for experimental analyses using jet binning \cite{Stewart:2011cf}.

\begin{figure}
\begin{center}
\includegraphics[width=8cm]{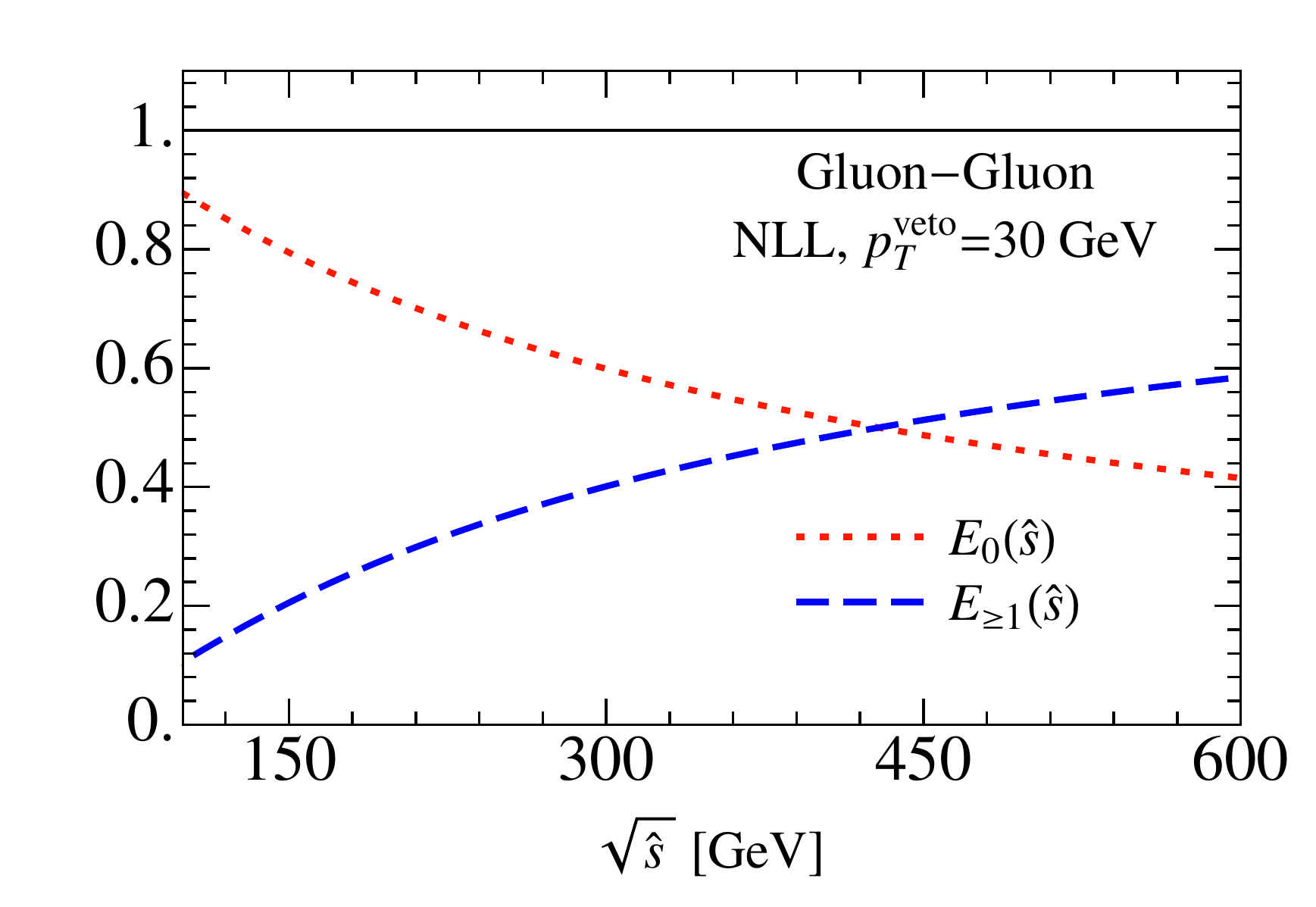}
\end{center}
\caption{The ratios $E_0(\hat s)$, $E_{\geq 1}(\hat s)$ for a gluon-gluon initial state, and $\ptveto=30$ GeV. There is a large migration from the exclusive 0-jet bin to the inclusive 1-jet bin as a function of $\hat s$. This phenomenon is important for understanding the impact of jet binning on off-shell cross sections.}
\label{fig:1jetincl}
\end{figure}

In \Fig{fig:1jetincl} we plot $E_0(\hat s)$, and
\begin{align} \label{eq:ratio_1jet}
E_{\geq 1}(\hat s)=\left .   \left (  \frac{  d\sigma_{\geq 1}^{NLL}(\ptveto)  } {d\sqrt{\hat s}}   \right )   \middle /      \left (  \frac{d\sigma}{d\sqrt{\hat s}} \right)      \right . ,
\end{align}
as a function of $\hat s $ for a gluon-gluon initial state with $\ptveto=30$ GeV. The behaviour in this plot is of course evident from \Fig{fig:lumi_ratio}, but it is interesting to interpret it in this fashion: as an $\hat s$ dependent migration between jet bins. Although our  calculation is only for the inclusive 1-jet bin, the dominant increase will be in the exclusive 1-jet bin.  

This migration between the jet bins as a function of $\hat s$ is important for the proper understanding of the off-shell cross section predictions in the presence of jet vetos. For CMS's recent off-shell $H\rightarrow ZZ\rightarrow 2l2\nu$ analysis, ignoring the VBF category, the events were categorized into exclusive zero jet, and inclusive one jet bins \cite{CMS:2014ala}, both of which have high sensitivity, due to the clean experimental signal. For the case of $H\rightarrow WW$, exclusive 0, 1, and 2 jet bins are used, although the experimental sensitivity is largest in the 0-jet bin, where the backgrounds are minimized. 

The effect of the migration is therefore different in the two cases. For $H\rightarrow ZZ$, since the backgrounds are easier to control, the jets that migrate from the exclusive 0-jet bin are captured in the inclusive 1-jet bin. Since both are used in the experiment, there is not a significant loss in analysis power. Accurate predictions for the two jet bins should still be used, and the correlations in the theory uncertainties due to resummation should still be treated properly.  For the case of $H\rightarrow WW$, where the jet veto plays a more essential role in removing backgrounds, the migration causes a loss in sensitivity. For example, the analysis of \cite{Campbell:2013wga} used the exclusive zero jet bin of $H\rightarrow WW$ to bound the Higgs width without a treatment of the $\hat s$ dependence induced by the jet veto. 
This will be discussed further in \Sec{sec:bound}. 
Calculations for the exclusive 1-jet and 2-jet bins are more difficult. Although NLL resummed results exist for the case $p_T^{jet} \sim \sqrt{\hat s}$ \cite{Liu:2013hba,Liu:2012sz}, the treatment of $p_T^{jet} \ll \sqrt{\hat s}$ is more involved~\cite{Boughezal:2013oha}.  The latter is the kinematic configuration of interest for far off-shell production.

\subsection{Variation with $E_\text{cm}$}\label{sec:Ecm}
Here we comment briefly on the dependence of the exclusive zero jet cross section on the hadronic centre of mass energy, $E_\text{cm}$. This is of course of interest as the LHC will resume at $E_\text{cm}=13$ TeV in the near future, and Higgs coupling and width measurements are important benchmarks for future colliders at higher energies. Here we only discuss the $\hat s$ dependence of the suppression due to the jet veto, the ratio of \Eq{eq:ratio}, on $E_\text{cm}$. Of course, with an increased $E_\text{cm}$, one can more easily achieve higher $\hat s$, allowing for off-shell production over a larger range, magnifying the importance of off-shell effects. We will discuss this for the specific case of $gg\rightarrow H \rightarrow WW$ in \Sec{sec:WW}.

\begin{figure}
\begin{center}
\includegraphics[width=8cm]{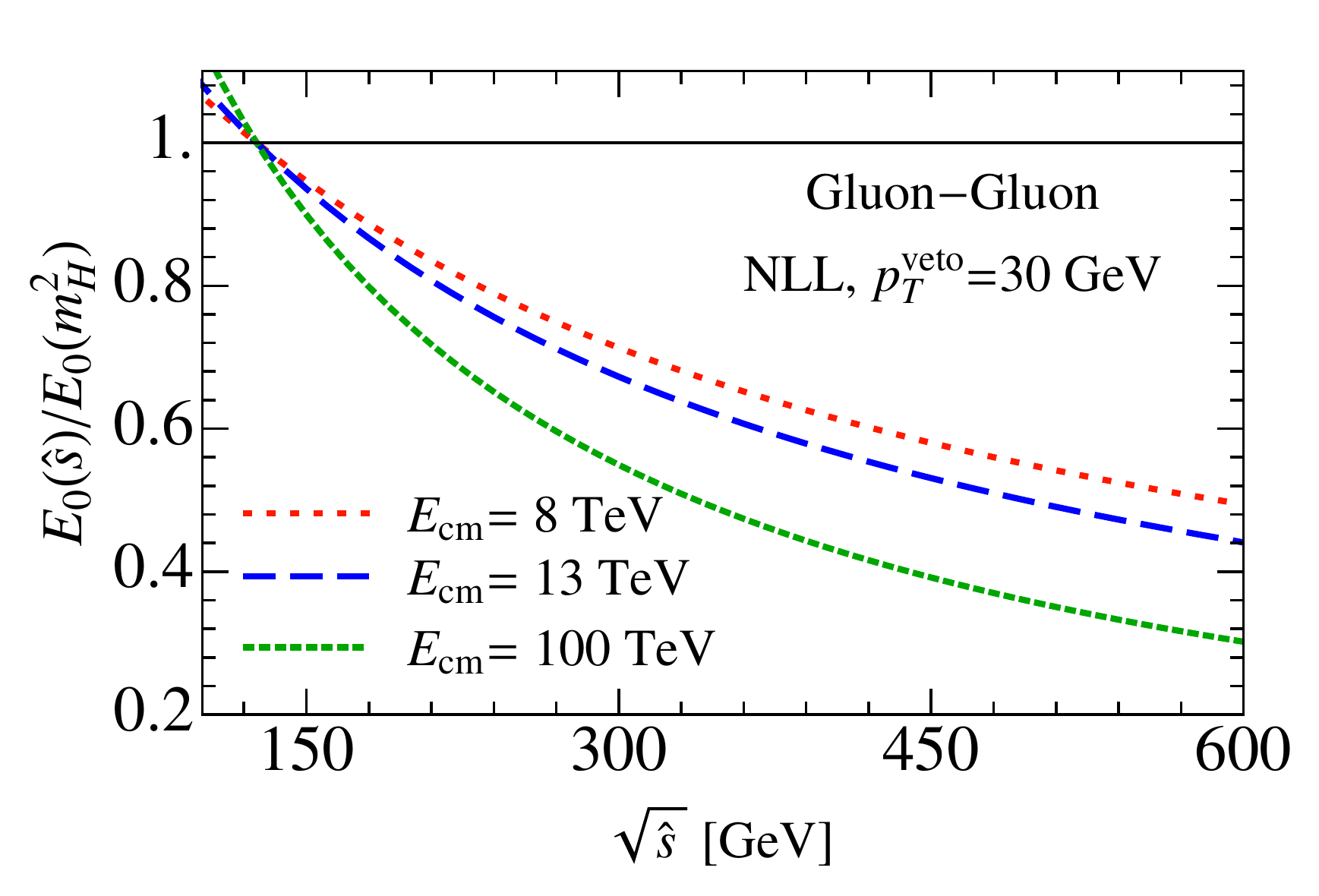}
\end{center}
\caption{A comparison of the effect of the jet veto at $E_\text{cm}=8,13,100$ TeV for a gluon-gluon initial state, and $\ptveto=30$ GeV. At higher $E_\text{cm}$ a larger suppression in the exclusive zero jet bin is observed, due to the larger range of Bjorken $x$ probed. }
\label{fig:Ecm}
\end{figure}

In \Fig{fig:Ecm} we compare the ratio $E_0(\hat s)/E_0(m^2_H)$ for $E_\text{cm}=8,13,100$ TeV. As the value of $E_{\text{cm}}$ is raised, the $\hat s $ dependence of the jet veto suppression systematically increases. Although the effect is relatively small between $8$ TeV and $13$ TeV,  it is significant at $100$ TeV. A similar effect was discussed in \cite{Campbell:2013qaa} where the exclusive zero jet fraction for on-shell Higgs production was observed to decrease with increasing $E_{\text{cm}}$. Since the Sudakov factor is independent of $E_\text{cm}$, this difference arises due to the fact that as the $E_{\text{cm}}$ is increased, the PDFs are probed over a larger range of Bjorken $x$, including smaller $x_{a,b}$ values. In the NLL factorization theorem of \Eq{eq:eqNLL} the PDFs are evaluated at the scale $\ptveto$ instead of at the scale $\hat s$. The impact of this change of scales in the PDFs depends on the $x$ values probed, and causes an increasing suppression as $E_{\text{cm}}$ is increased. 

For the majority of this paper we will restrict ourselves to $E_{\text{cm}}=8$TeV, although in \Sec{sec:higherEcm} we will further discuss the effect of an increased $E_{\text{cm}}$.

\section{$gg\rightarrow H\rightarrow WW$: A Case Study}\label{sec:WW}

In this section we use $gg\rightarrow H\rightarrow WW$ to discuss the effect of an exclusive jet veto in more detail. $H\rightarrow WW$ is a particularly interesting example to demonstrate the $\sqrt{\hat s}$ dependence of the jet veto suppression since it has a sizeable contribution from far off-shell production \cite{Uhlemann:2008pm,Kauer:2013cga}, and furthermore has interference with continuum $gg\rightarrow WW\rightarrow l\nu l\nu$ production, which contributes over a large range of $\sqrt{\hat s}$ \cite{Campbell:2011cu,Kauer:2012ma,Kauer:2013qba}. A jet veto is also required experimentally for this channel due to large backgrounds. For the signal-background interference, we will consider two different Higgs masses, $m_H=126$ GeV and $m_H=600$ GeV, which have interference which depend differently on $\sqrt{\hat s}$ and therefore cover two interesting scenarios for the different effects that the jet veto can have.

In \Sec{sec:hardfunc} we discuss in detail the hard coefficients, and the matching to SCET.  Default parameters are given in \Sec{sec:nums}.  In \Sec{sec:conv} we use $gg\rightarrow H\rightarrow WW \rightarrow l\nu l\nu$, which can be calculated to NLL and NNLL, to study the convergence in the off-shell region. The extension to NNLL allows us to study the effect of the finite radius of the jet veto. In \Sec{sec:int_126} we show results for the NLL resummation for the signal-background interference. Although we are unable to go to NNLL without the NLO hard function for the interference, the results of \Sec{sec:conv} give us confidence that the NLL result is capturing the dominant effects imposed by the jet veto restriction. In \Sec{sec:mt_126} we consider jet veto suppression in the exclusive zero jet bin as a function of the experimental observable $M_T$.

\subsection{Hard Function and Matching to SCET}\label{sec:hardfunc}
In this section we discuss the hard function appearing in the SCET factorization theorem, which carries the dependence on the hard underlying process. This is discussed in some detail, as we will be considering signal-background interference, which has not previously been discussed in the language of SCET.

\begin{figure}
\begin{center}
\subfloat[]{
\includegraphics[width=6.0cm]{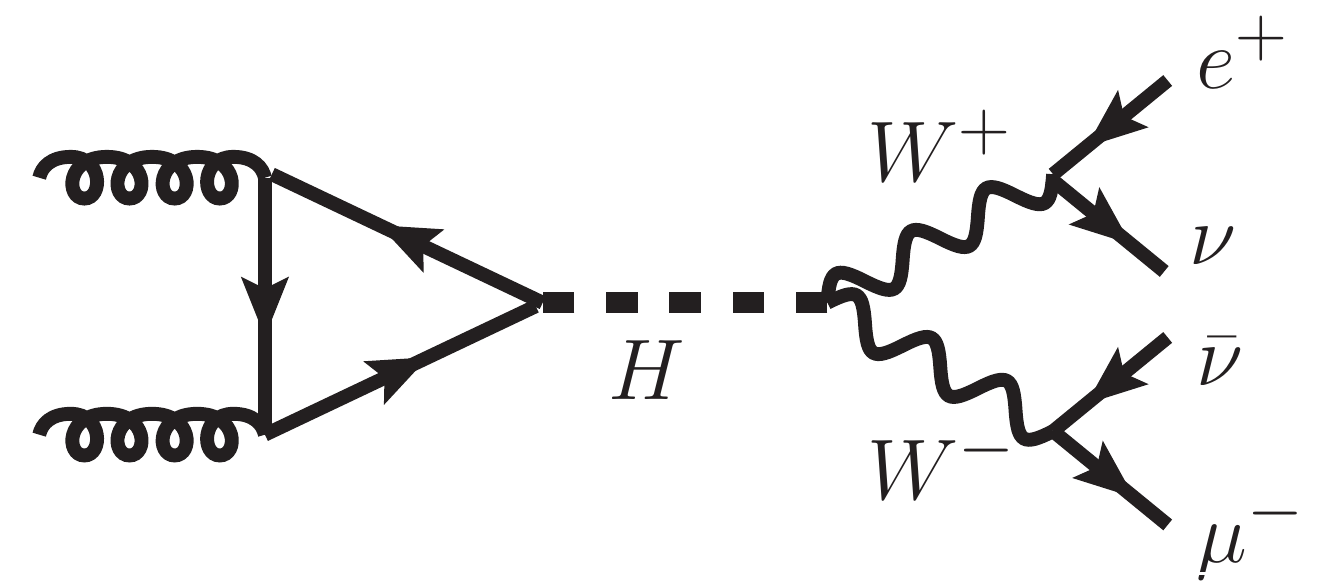}
}
$\qquad$
\subfloat[]{
\includegraphics[width=5.0cm]{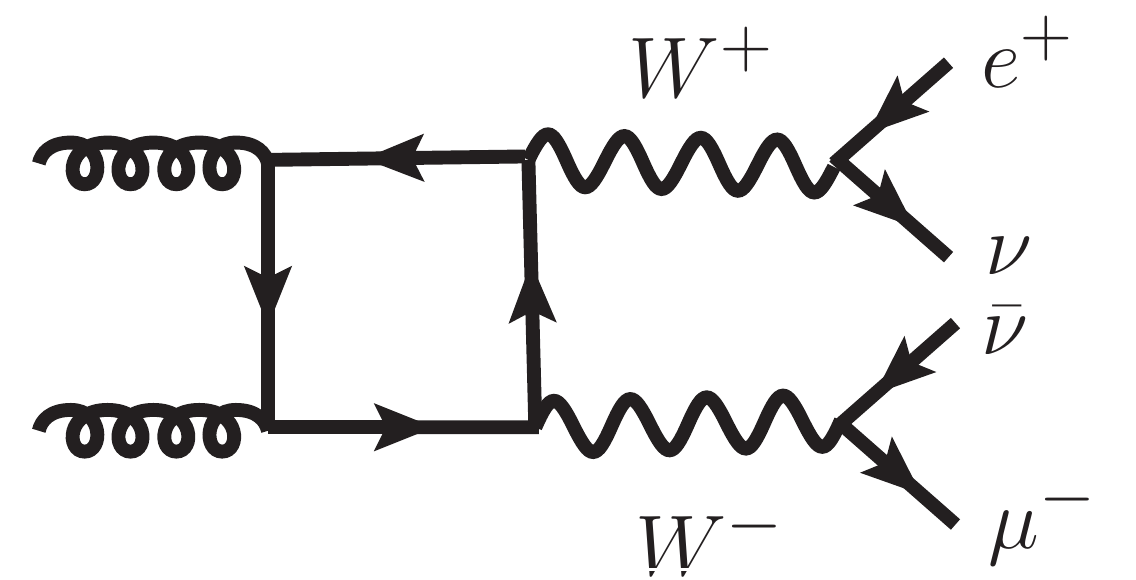}
}
\end{center}
\caption{LO Higgs mediated, (a), and continuum, (b), diagrams contributing to the process $gg\rightarrow l\nu l\nu$. These are matched onto the helicity basis of SCET operators given in \Eq{eq:operators}.}
 \label{fig:diagrams}
\end{figure}

It was shown in \cite{Campbell:2011cu} that only two Feynman diagram topologies contribute to the process $gg\rightarrow \nu_e e^+ \mu^- \bar \nu_\mu$ at LO, due to a cancellation between diagrams with an s-channel Z boson. The two diagrams that contribute are the gluon-fusion Higgs diagram, and a quark box diagram for the continuum production, both of which are shown in \Fig{fig:diagrams}. The $gg\rightarrow  \nu_e e^+ \mu^- \bar \nu_\mu$ cross section consists of Higgs production, the continuum production, and the interference between the two diagrams. Although the interference contribution is small when considering on-shell Higgs production, it becomes important in the off-shell region.  

In the effective field theory formalism, these two diagrams are matched onto effective operators in SCET. It is convenient both for understanding the interference, and for comparing with fixed order QCD calculations to work in a helicity and color operator basis in SCET \cite{Stewart:2012yh,Stewart:forthcoming}. For this process the color structure is unique, as we are considering the production of a color singlet state from two gluons. We therefore focus on the helicity structure. The helicity of the outgoing leptons is fixed by the structure of the weak interactions, so we need only construct a helicity basis for the incoming gluons. We write the amplitudes for the above diagrams as
\be
\cA_H(1_g^{h_1},2_g^{h_2},3_{\nu_e}^-,4_{\bar e}^+,5_{\mu}^-,6_{\bar \nu_\mu}^+),~~\cA_\cC(1_g^{h_1},2_g^{h_2},3_{\nu_e}^-,4_{\bar e}^+,5_{\mu}^-,6_{\bar \nu_\mu}^+)
\ee  
where the subscripts $H$, $\cC$ denote the Higgs mediated, and continuum box mediated diagrams respectively, and the superscripts denote helicity. In the following we will mostly suppress the lepton arguments, as their helicities are fixed, and focus on the gluon helicities. 

Since the SM Higgs boson is a scalar, we have
\be
\cA_H(1_g^{-},2_g^{+})=\cA_H(1_g^{+},2_g^{-})=0.
\ee
In this paper, our focus is on the Higgs production and the signal-background interference. Since there is no interference between distinct helicity configurations, we can therefore also ignore the continuum production diagrams with the $+-$, $-+$ helicity configuration. These do contribute to the background, however their contribution is small compared to the $q\bar q \rightarrow l\nu l\nu$ process. 

The above amplitudes are matched onto operators in the effective theory. The SCET operators at leading power are constructed from collinear gauge-invariant gluon fields \cite{Bauer:2000yr,Bauer:2001ct}
\be
\cB^\mu_{n,\omega \perp}= \frac{1}{g} \left[ \delta(\omega +\bar {\cP}_n) W_n^\dagger(x) i \cD^\mu_{n\perp} W_n(x)   \right]
\ee
where $n$, $\bar n$ are lightlike vectors along the beamline. The collinear covariant derivative is defined as
\be
i \cD^\mu_{n\perp} =\cP^\mu_{n\perp}+gA^\mu_{n\perp},
\ee
with $\cP$ a label operator which extracts the label component of the momentum in the effective theory, and $W_n$ is a Wilson line defined by
\be
W_n(x)= \left [   \sum\limits_{\text{perms}} \exp \left (  -\frac{g}{\bar \cP_n} \bar n \cdot A_n(x)  \right )    \right ].
\ee
A  helicity basis of SCET operators for the process of interest is given by
\begin{align}\label{eq:operators}
&\cO^{++}=\frac{1}{2}\cB^a_{n+}\cB^a_{\bar n+} J_{34-}J_{56-}\\
&\cO^{--}=\frac{1}{2}\cB^a_{n-}\cB^a_{\bar n-}   J_{34-}J_{56-},
\end{align}
where the $1/2$ is a bosonic symmetry factor to simplify matching to the effective theory. We have defined collinear gluon fields of definite helicity by
\be
\cB^a_{i \pm} =-\epsilon_{\mp \mu} \cB^{a,\mu}_{n_i,\omega_i \perp_i},
\ee
where $\epsilon_{\mp \mu}$ are polarization vectors, as well as leptonic currents of definite helicity
\be
J_{ij-}=\epsilon_{+}^\mu(p_i,p_j)\frac{\bar \psi_{i-}\psi_{j-}}{\sqrt{2}[ji]}.
\ee
In this expression, and in the expressions for the Wilson coefficients given below, we will use the standard spinor helicity notation, with $\langle ij \rangle= \bar u_-(p_i) u_+(p_j)$, and $[ ij ]= \bar u_+(p_i) u_-(p_j)$.

Note that we use Hard functions that are fully differential in the leptonic momenta. This allows for realistic experimental cuts on the leptonic phase space to be straightforwardly incorporated.

It is important to note that operators with distinct external helicities do not mix under the SCET RGE at leading power. The jets from the incoming partons, which are described by the beam functions, can only exchange soft gluons, described by the soft function. At leading power, the soft gluons cannot exchange spin, only color, and therefore the RGE can only mix Wilson coefficients in color space, which in this case is trivial.  This allows one to consistently neglect the operators $\cO^{+-}$, $\cO^{-+}$, which would arise from matching the $\cA_{\cC}(1_g^{-},2_g^{+})$, and $\cA_{\cC}(1_g^{+},2_g^{-})$ onto SCET. They do not contribute to the process of interest, and do not mix under the RGE with the operators that do contribute. 

We are interested in considering both the direct Higgs production and signal-background interference separately, so it is convenient to maintain this distinction in SCET. Although the SCET operators are the same in both cases, we can separate the Wilson coefficient into a component from the Higgs mediated diagram, and a component from the box mediated continuum diagram. We then have four Wilson coefficients
\be
C^H_{++},~C^H_{--},~C^C_{++},~C^C_{--}  \,.
\ee
Since the operators are in a helicity basis, these four Wilson coefficients are simply the finite part of the helicity amplitudes for the given processes (or more specifically for ${\overline{\rm MS}}$ Wilson coefficients in SCET are the finite part of the helicity amplitudes computed in pure dimensional regularization). These were computed in \cite{Campbell:2011cu}, and can be obtained from the MCFM code \cite{Campbell:2010ff}. The Wilson coefficients for the Higgs mediated process depend on the Higgs and W boson widths and masses, as well as the invariants $s_{12}, s_{34}, s_{56}$. The explicit leading order Wilson coefficients for the Higgs mediated process are given by
\begin{align}
&C^H_{--}(m_H,\Gamma_H, s_{12}, s_{34}, s_{56})= \left( \frac{g_w^4 g_s^2}{16\pi^2} \right) \cP_H(s_{12})   \cP_W(s_{34})    \cP_W(s_{56}) \frac{\langle 12 \rangle \langle 35 \rangle [64]}{[21] s_{34} s_{56}} F_H(s_{12}) , \\
&C^H_{++}(m_H,\Gamma_H, s_{12}, s_{34}, s_{56})= \left( \frac{g_w^4 g_s^2}{16\pi^2} \right) \cP_H(s_{12})   \cP_W(s_{34})    \cP_W(s_{56}) \frac{[ 12 ] \langle 35 \rangle [64]}{\langle 21\rangle s_{34} s_{56}} F_H(s_{12}),
\end{align}
where the function $\cP_i$ is the ratio of the propagator for the particle species $i$ to that of the photon,
\be
\cP_i(s)=\frac{s}{s-m_i^2+i\Gamma_i m_i}   \,.
\ee
We have also used $F_H (s_{12})$ for the usual loop function for gluon-fusion Higgs production
\be
F_H(s_{12})=\sum\limits_{q=t,b} \frac{m_q^2}{s_{12}}\bigg[   2+\Big(  \frac{4m_q^2}{s_{12}}-1  \Big) g\Big(\frac{m_q^2}{s_{12}} \Big)   \bigg] \,,
\ee
with
\be
g(x)= \begin{cases} 
      \frac{1}{2} \left[ \log \left(  \frac{1+\sqrt{1-4x}}{1-\sqrt{1-4x}}   \right ) -i\pi   \right]^2 & x< \frac{1}{4} \\
      -2 \left ( \sin^{-1} \left (  \frac{1}{2\sqrt{x}}\right)   \right )^2 & x\geq \frac{1}{4}. 
   \end{cases}
\ee

The Wilson coefficients $C_{++}^C$ and $C_{--}^C$ for the box diagram depend on the W mass and width, as well as the kinematic invariants formed by the external momenta. In the presence of massive quarks in the loops, they are extremely lengthy, so we do not reproduce them here. We refer interested readers to \cite{Campbell:2011cu}, and the MCFM code from which we have extracted the required results for our analysis. We have verified that our extracted expressions reproduce quoted numerical results and distributions in \cite{Campbell:2011cu}.

The Hard coefficient, $H$, appearing in the factorization theorem, \Eq{eq:fact} is given by the square of the Wilson coefficients:
\be
H=|C_{++}^H+C_{++}^C|^2 + |C_{--}^H+C_{--}^C|^2  + | C_{+-}^C |^2 + | C_{-+}^C |^2 
\ee
As is typically done in the case of squared matrix elements, we can separate the hard function into the sum of a hard function for the Higgs mediated process $H^H$, a hard function for the interference $H^{int}$, and a hard function for the background arising as the square of Wilson coefficient for the continuum process $H^C$ (which we will not use here). For the first two we have
\begin{align}
&H^H=|C_{++}^H|^2+|C_{--}^H|^2\\
&H^{int}=2\text{Re}\left [C_{++}^H (C_{++}^C)^\dagger\right ]   +   2\text{Re}\left [C_{--}^H (C_{--}^C)^\dagger\right ].
\end{align}
This decomposition allows us to discuss the resummation of the interference and the Higgs mediated processes separately in the effective theory, in a language that is identical to that used in Feynman diagram calculations. In \Secs{sec:conv}{sec:int_126} we will discuss the effect of resummation on both the Higgs mediated contribution and the signal-background interference.

\subsection{Parameters for Numerical Calculations} \label{sec:nums}

For the numerical results, we use the default set of electroweak parameters from MCFM, following \cite{Campbell:2013wga,Campbell:2011cu}:
\begin{align*}
&m_W=80.398\,\text{GeV},  &m_Z&=91.1876\,\text{GeV} \,, \\
&\Gamma_W=2.1054\,\text{GeV},  &\Gamma_Z& =2.4952\,\text{GeV}\,, \\
&m_t=172.5\,\text{GeV}, &m_b&=4.4\,\text{GeV}\,, \\
&G_F=1.16639 \times 10^{-5}\, \text{GeV}^{-2},  &\sin^2 \theta_W &=0.222646\,, \\
&\alpha_{e.m.}(m_Z)=\frac{1}{132.338} \,.
\end{align*}
We use the following two Higgs mass/width combinations to demonstrate the dependence on the Higgs mass \footnote{Although more recent analyses point to a Higgs mass closer to $125$ GeV, we have taken $126$ GeV as representative of a light Higgs. The conclusions of this section do not depend on this small difference for the Higgs mass, and our plots for $125$ and $126$ GeV are indistinguishable at the resolution shown. }: 
\begin{align*}
&m_H=126\,{\rm GeV},\qquad \Gamma_H=0.004307 \,\text{GeV} \,,\\
&m_H=600\,{\rm GeV},\qquad \Gamma_H=122.5\,\text{GeV}  \,,
\end{align*}
where the widths are determined from HDECAY~\cite{Djouadi:1997yw}.
We use the NLO PDF fit of Martin, Stirling, Thorne and Watt \cite{Martin:2009iq} with $\alpha_s(m_Z)=0.12018$. 

The results in this section were obtained using the analytic results for the partonic process documented in \cite{Campbell:2011cu}. Scalar loop integrals were evaluated using the LoopTools package \cite{Hahn:2000jm}, and phase space integrals were done using the Cuba integration package \cite{Hahn:2004fe}. For all the results presented in this section, we have integrated over the leptonic phase space, and allow for off-shell vector bosons.

\subsection{Off-Shell Higgs Production} \label{sec:conv}

We begin by studying the effect of the jet veto on far off-shell Higgs production in $gg\rightarrow H \rightarrow WW\rightarrow e^+\nu_e \mu \bar \nu_\mu$. While a full analysis of the off-shell region also requires the inclusion of signal-background interference, for which the hard function to NLO is not known, we use the off-shell Higgs mediated process to study the convergence of the resummed predictions. In particular, one is first sensitive to the jet radius at NNLL. The ability to perform the resummation to NNLL for the Higgs mediated signal enables us to assess the convergence of the resummed predictions in the off-shell region. It also allows us to check that the NLL result, which will be used when signal-background interference is included, accurately captures the effect of the jet veto reasonably well. In particular, we will focus on the shape of the differential distribution in $\hat s$. As will be discussed in more detail in \Sec{sec:bound}, the procedure for extracting a bound on the Higgs width from the off-shell cross section uses a rescaling procedure to the on-shell cross section. Because of this, the shape, but not the normalization of the distribution is important for an accurate application of this method. Therefore, as in \Sec{general} we will rescale the differential cross sections by $E_0(m_H^2)$, allowing us to focus just on the shape.
\begin{figure}
\begin{center}
\subfloat[]{\label{fig:conv_30}
\includegraphics[width=6cm]{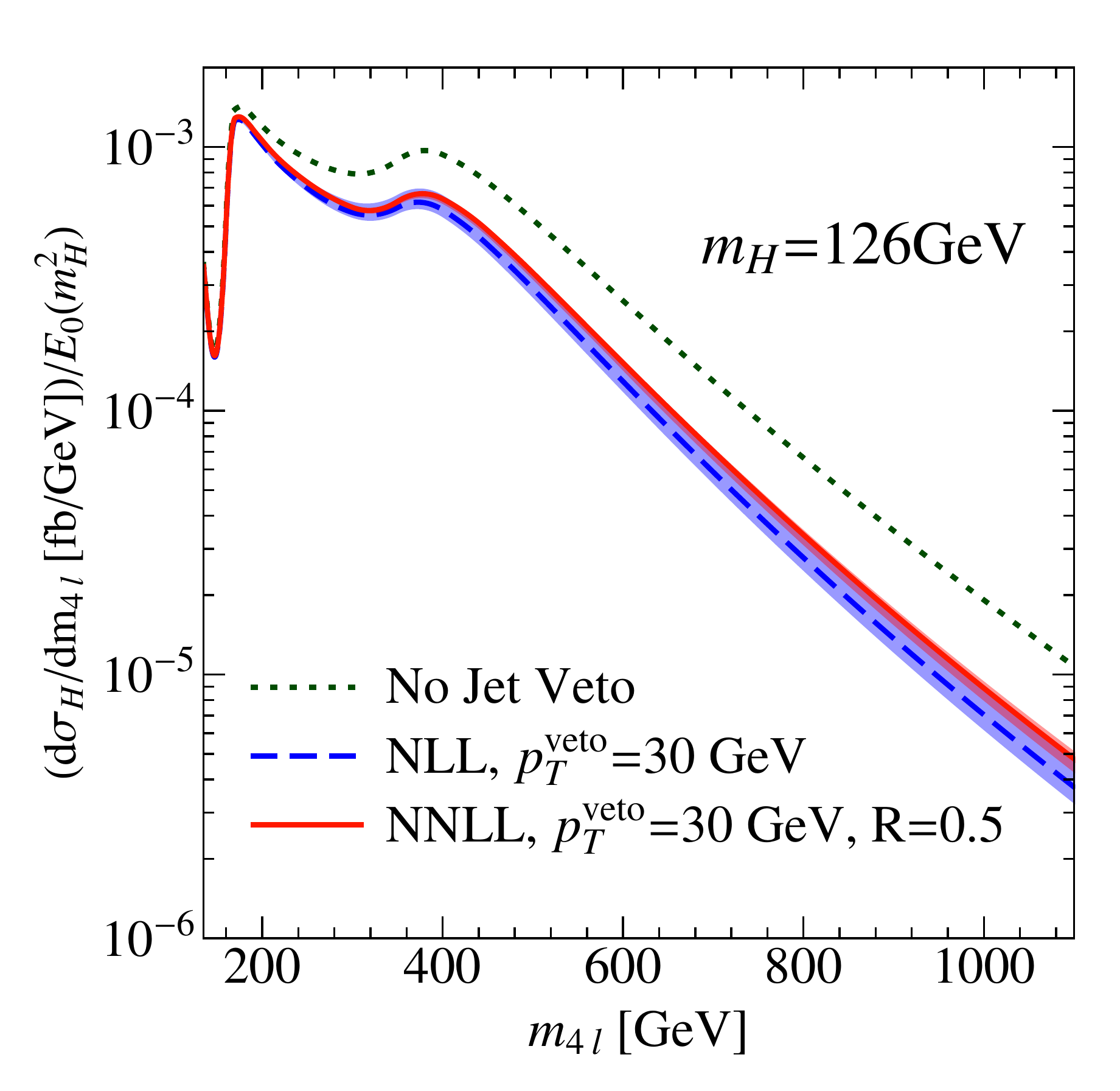}
}
$\qquad$
\subfloat[]{\label{fig:conv_20}
\includegraphics[width=6cm]{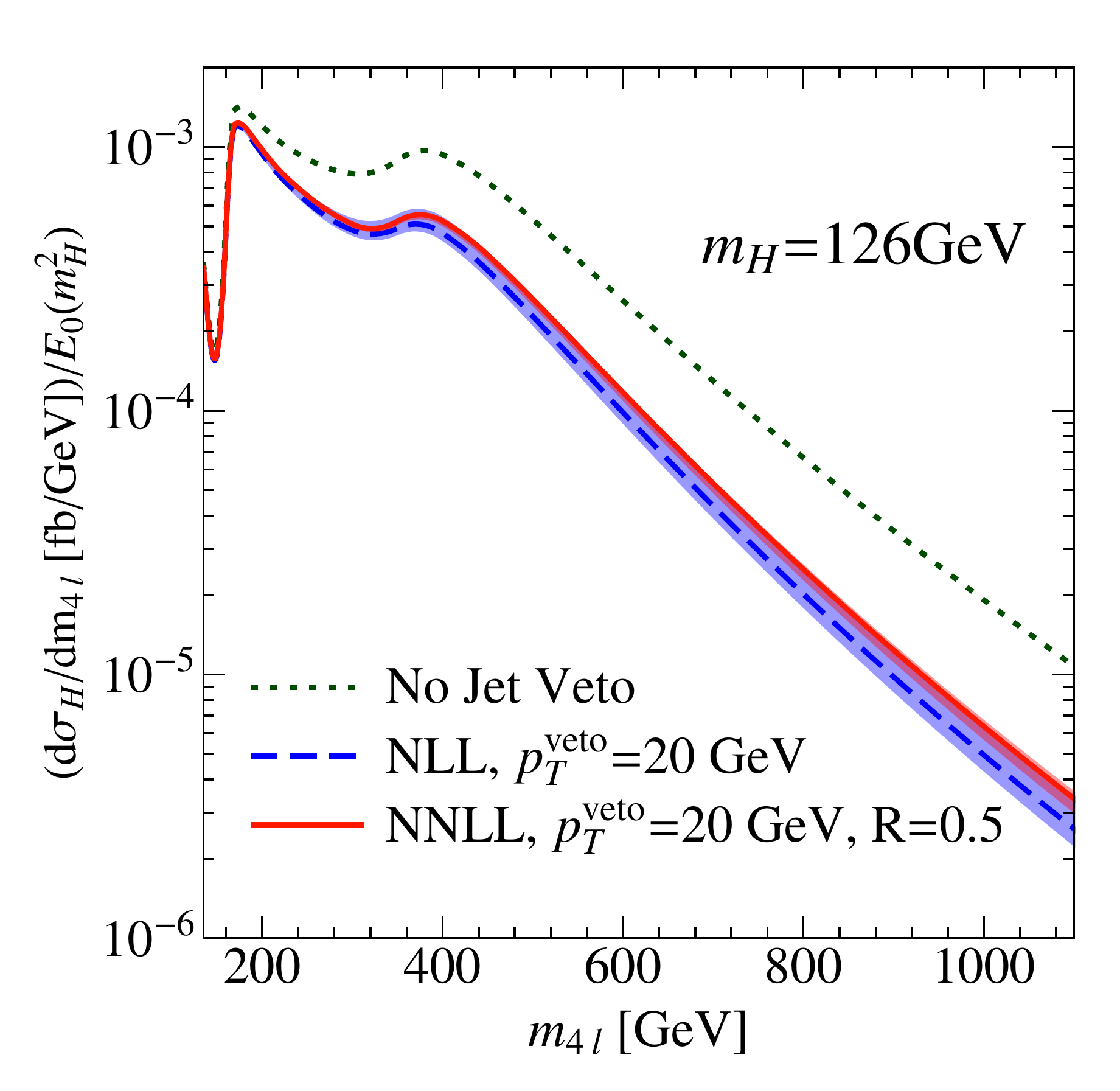}
}
\end{center}
\caption{The off-shell Higgs cross section in the exclusive zero jet bin for $\ptveto=30$ GeV in (a), and $\ptveto=20$ GeV in (b), with $R=0.5$ in both cases. Results are normalized by the jet veto suppression at the Higgs mass, such that the on-shell cross section is the same in all cases, allowing one to focus on the modification to the shape of the distribution. NLL and NNLL results are similar, with a small modification due to the finite jet radius, which is not present in the NLL calculation. }
\label{fig:conv}
\end{figure}

The NNLL calculation requires the NNLL beam and soft functions, which are known in the literature for a jet veto defined by a cut on $p_T$ \cite{Stewart:2013faa}, as well as the virtual part of the NLO gluon-fusion hard function. The NLO virtual contributions for gluon fusion Higgs production are known analytically with full dependence on the top and bottom quark mass \cite{Harlander:2005rq}, which is necessary, as in the off-shell region one transitions through the $\sqrt{\hat s} =2m_t$ threshold.\footnote{The analytic NLO virtual corrections were also used in \cite{Banfi:2013eda} to study the dependence of the jet veto on the b-quark mass for the case of on-shell Higgs production.} The NLO hard function is determined by matching onto the gluon-fusion operators in SCET, as discussed in \Sec{sec:hardfunc}. We do not include in our calculation the non-singular pieces, as we focus on the region $\ptveto \ll \sqrt{\hat s}$, where the singular contributions dominate, and are not interested in the transition to the region $\ptveto \sim \sqrt{\hat s}$.

In \Fig{fig:conv} we plot the resummed distribution, normalized to the jet veto suppression at the Higgs mass: $(d\sigma_0/dm_{4l})/E_0(m_H^2)$, for off-shell $gg \rightarrow H \rightarrow WW \rightarrow e^+\nu_e \mu \bar \nu_\mu$. Note that in the case without a jet veto, the jet veto suppression at the Higgs mass is defined to be $1$. We have integrated over the leptonic phase space. Here $m_{4l}=\sqrt{\hat s}$ is the invariant mass of the 4 lepton final state. In \Fig{fig:conv_30} we use $\ptveto=30$ GeV, and in \Fig{fig:conv_20} we use $\ptveto=20$ GeV. In both cases, we use a jet radius of $R=0.5$, as is currently used by the CMS collaboration. The uncertainty bands are rough uncertainty estimates from scale variations by a factor of 2. Note that in the calculation, we use a five flavour scheme, even above $m_t$ since the difference with using a six flavour coupling is well within our error band.

\Figs{fig:conv_30}{fig:conv_20} show a small modification to the differential distribution between NLL and NNLL. This arises primarily due to the clustering logarithms, which introduce dependence on the jet radius, which is not present at NLL. The $R$ dependence reproduces the expected physical dependence of the cross section on $R$: for a fixed $\ptveto$ cut, the restriction on radiation from the initial partons becomes weaker as the jet radius is decreased, causing a smaller suppression of the cross section. Despite this, the shape is well described by the NLL result.  In particular, the NLL result captures the dominant effect of the exclusive jet veto on the off-shell cross section. This is important for the resummation of the interference, considered in \Sec{sec:int_126}. In this case, higher order results are not available (for some approximate results, see \cite{Bonvini:2013jha}), and therefore one is restricted to an NLL resummation. However, the results of this section demonstrate that the NLL result accurately captures the effects of the jet veto on the shape of the distribution as a function of $\hat s$.

\subsection{Signal-Background Interference}\label{sec:int_126}
Signal-background interference for the process $gg\rightarrow  l\nu l\nu$ has been well studied in the literature\cite{Campbell:2011cu,Kauer:2012ma,Kauer:2013qba}. The interference comes almost exclusively from the $\sqrt{\hat s}>2M_W$ region. For a light Higgs, $m_H< 2M_W$, this means that the interference comes entirely from $\sqrt{\hat s} > m_H$. For a heavy Higgs, the Higgs width is sufficiently large that there are contributions to the interference from a wide range of $\sqrt{\hat s}$. The signal-background interference is therefore, in both cases, an interesting process on which to demonstrate the effect of the jet veto.

The NLO virtual corrections are not available for the interference process, restricting the resummation accuracy to NLL. However, as argued in \Sec{sec:conv} if one is interested in the shape of the distribution, and not the normalization, the NLL captures the effects of the jet veto. One thing that cannot be known without a full calculation of the NLO virtual contributions to the interference is if the NLO virtual contributions for the interference are different than for the signal. For the case of interference in $H\rightarrow \gamma \gamma$ where they are known, the virtual contributions for the interference were found to be smaller than for the signal \cite{Dixon:2013haa}. Due to the similar structure of the diagrams for $H\rightarrow WW$, the same could certainly be true. However, we expect this to be a minor correction compared to the effects of the jet veto. In particular, we do not expect the K-factor to have strong $\hat s$ dependence, which is the important effect captured by the resummation. In this section we use the LO result for $gg \rightarrow e\nu \mu \nu$, fully differential in the lepton momenta, which is available in the MCFM code, and is documented in \cite{Campbell:2011cu}.

We begin by reviewing the notation for the signal-background interference in $gg\rightarrow e \nu \mu \nu$ at LO following \cite{Campbell:2011cu}. It is convenient to pull out the dependence on $m_H$ and $\Gamma_H$ coming from the s-channel Higgs propagator. Defining $\widetilde C^H=(\hat s-m_H^2+im_H \Gamma_H)C^H$, we can separate the Hard function for the signal-background interference into its so called ``Imaginary'' and ``Real'' contributions:
\be \label{eq:int_decomp}
H^{int}=\frac{2(\hat s-m_H^2)}{(\hat s-m_H^2)^2+m_H^2 \Gamma_H^2} \text{Re}\left [ \widetilde C^H( C^C)^\dagger \right ]+\frac{2m_H \Gamma_H}{(\hat s-m_H^2)^2+m_H^2 \Gamma_H^2}\text{Im}\left [ \widetilde C^H (C^C)^\dagger \right ].
\ee
In \Eq{eq:int_decomp} there is a sum over helicities of the Wilson coefficients, which for notational convenience has not been made explicit. Note that the imaginary part of the interference is multiplied by an explicit factor of $\Gamma_H$, and is therefore negligible for a light Higgs. 

\begin{figure}
\begin{center}
\subfloat[]{\label{fig:int_126}
\includegraphics[width=7cm]{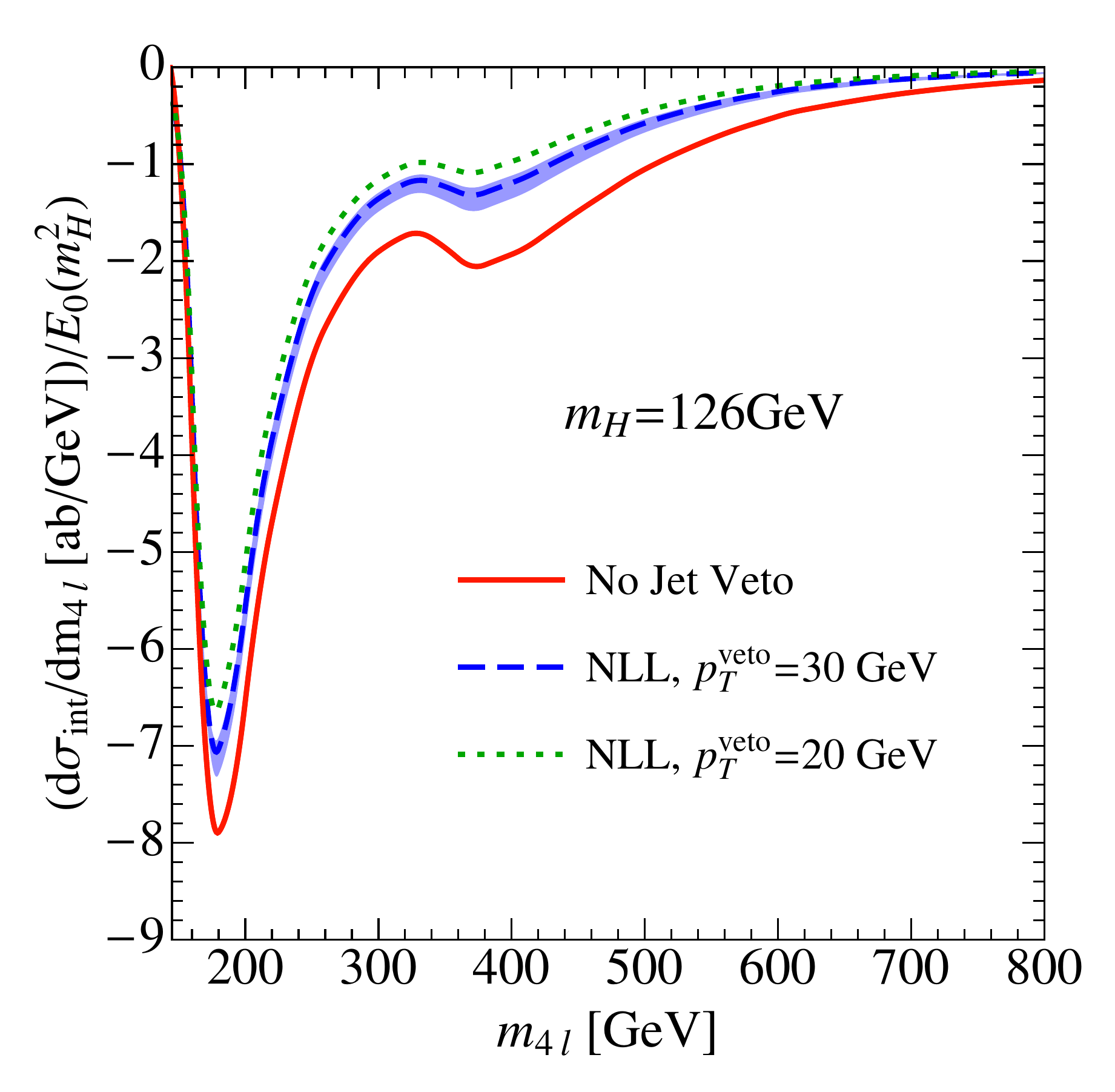} 
}
$\qquad$
\subfloat[]{\label{fig:int_600}
\includegraphics[width=6.9cm]{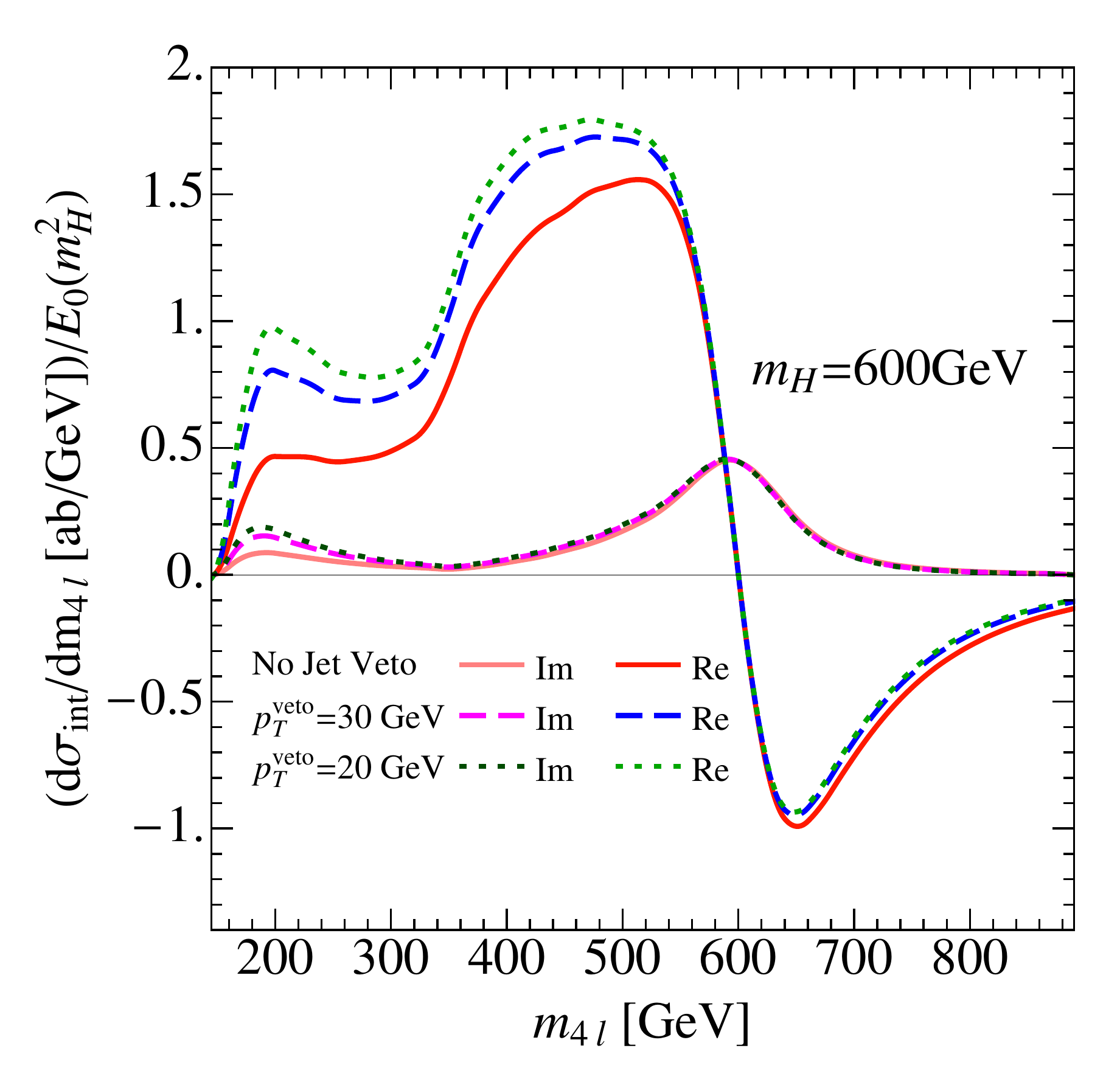}
}
\end{center}
\caption{Signal-background interference in $gg\rightarrow e^+ \nu_e \mu \bar \nu_\mu$ for (a) $m_H=126$ GeV, and (b) $m_H=600$ GeV. NLL predictions are shown for $\ptveto=20,30$ GeV, and have been rescaled by the jet veto efficiency at $m_H$. The size of the signal-background interference relative to the on-shell cross section is enhanced by the jet veto for a heavy Higgs, whereas it is suppressed for a light Higgs.  For $m_H=126$ GeV the jet veto causes a significant reduction of the cross section in the far off-shell region relative to the on-shell cross section.}
\label{fig:int}
\end{figure}

The interference without a jet veto is shown in \Fig{fig:int_126} for a $126$ GeV Higgs and \Fig{fig:int_600} for a $600$ GeV Higgs, as a function of $m_{4l}$. We have integrated over the phase space of the leptons, including allowing for off-shell vector bosons. The interference is negligible below the $\sqrt{\hat s}=2m_W$ threshold. For the case of $m_H=126$ GeV the only non-negligible contribution is the real part of the interference above the Higgs pole, which gives a negative contribution to the total cross section. In the case of $m_H=600$ GeV, there is significant interference both above and below the Higgs pole, and from both the real and imaginary parts. The interference below the pole dominates, leading to a net positive contribution to the total cross section. We have chosen these two Higgs masses, where the interference has a different $\sqrt{\hat s}$ dependence, so as to demonstrate the different effects that a jet veto can have on signal-background interference.

\Fig{fig:int} also shows as a function of $\sqrt{\hat s}$ the result for interference including a jet veto of $\ptveto=20,30$ GeV with NLL resummation, which can be compared with the interference without a jet veto. To make the interpretation of \Fig{fig:int} as simple as possible, we have rescaled the interference by $E_0(m^2_H)$, the jet veto efficiency at $m_H$. Therefore, enhancements and suppressions in the jet vetoed interference correspond to enhancements and suppressions of the interference relative to the on-shell Higgs contribution when a jet veto is applied. As expected from the discussion in \Sec{general}, we find a significant suppression of the interference at higher $\sqrt{\hat s}$, and this suppression increases with $\sqrt{\hat s}$.
 For $m_H=126$ GeV, shown in \Fig{fig:int_126}, the interference comes entirely from above $\sqrt{\hat s}=m_H$, and is therefore more highly suppressed by the jet veto relative to the on-shell Higgs cross section. 
 However, the situation is quite different for  the case of $m_H=600$ GeV, shown in \Fig{fig:int_600}.  Here the dominant contribution to the interference is from the real part in \eq{int_decomp}, which changes sign at $\sqrt{\hat s}=m_H$. The real part of the interference coming from below $\sqrt{\hat s}=m_H$ is positive and is partly cancelled by negative interference from above $\sqrt{\hat s}=m_H$ if we integrate over $\hat s$. The jet veto suppresses the on-shell cross section and the negative interference from above $\sqrt{\hat s}=m_H$ more than the contribution from the positive interference below $\sqrt{\hat s}=m_H$, and therefore the jet veto acts to enhance the interference contribution relative to the signal. This enhancement is significant in the case of $m_H=600$ GeV, as the interference has contributions starting at $m_{4l} \simeq 2m_W$, where the suppression due to the jet veto is smaller. To quantify this further we can consider the effect of the jet veto on the ratio
\be\label{eq:int_ratio}
R_I =\frac{\sigma_{H+I}}{\sigma_{H}} \,,
\ee
where $\sigma_{H+I}$ is the cross section including the signal-background interference, and $\sigma_H$ is the Higgs mediated cross section. The behaviour of this ratio is different for the two Higgs masses considered. Numerical values of $R_I$ are shown in \Tab{tab:ratio}. The effect of interference for $m_H=126$ GeV with or without the jet veto is fairly small, and would be made even smaller when cuts are made to eliminate interference.  However, for $m_H=126$ GeV the effect of the jet-veto can also be significantly amplified when cuts are used to maximize sensitivity to the Higgs width. For example, the analysis of \cite{Campbell:2013wga} considered the region $M_T>300$ GeV to bound the Higgs width. Since $m_{4l} \geq M_T$, we see from \Fig{fig:int_126} that in this region the effect of the exclusive jet veto is by no means a small effect, giving a suppression of $\sim 1.5- 2$. A representative error band from scale variation is also shown in \Fig{fig:int_126}. The effect on the derived bound will be discussed in \Sec{sec:bound}.

These two examples demonstrate that a jet veto can have an interesting interplay with signal-background interference, enhancing or suppressing its contribution relative to the Higgs mediated cross section, depending on the particular form of the interference. A detailed understanding of the interference is of phenomenological interest for both a light and heavy Higgs. In the case of $m_H=126$ GeV, the interference can be efficiently removed by cuts when studying the on-shell cross section \cite{Campbell:2011cu}, but is important for the understanding of the off-shell cross section. In the case of a heavy Higgs, the interference is important for heavy Higgs searches \cite{Campbell:2011cu,TheATLAScollaboration:2013zha,Chatrchyan:2013yoa}, where it is a large effect, and cannot be easily removed by cuts. The effect of the jet veto must therefore be incorporated in such searches.
\begin{table}[t]
\begin{center}
\begin{tabular}{c|c|c|c}
&No Veto & $\ptveto=30$ GeV & $\ptveto=20$ GeV \\ 
\hline
$m_H=126$ GeV & 0.92 & 0.94 &  0.95\\
$m_H=600$ GeV &1.38& 1.49 &  1.54 \\
\end{tabular}
\end{center}
\caption{Values of $R_I=\frac{\sigma_{H+I}}{\sigma_{H}}$ for $m_H=126$ GeV and $m_H=600$ GeV for two different values of $\ptveto$. As is clear from \Fig{fig:int}, the jet veto causes a suppression of the importance of the interference relative to the Higgs mediated process for a light Higgs, and an enhancement for a heavy Higgs. }
\label{tab:ratio}
\end{table}

\subsection{Suppression as a Function of $M_T$}\label{sec:mt_126}

We have so far discussed the effect of the jet veto on the cross section as a function of $\sqrt{\hat s}$, as the Sudakov factor is explicitly a function of $\sqrt{\hat s}$. However, in the case of $H\rightarrow WW\rightarrow l \nu l\nu$, the total invariant mass of the leptons cannot be reconstructed. A substitute for $\sqrt{\hat s}$, used in \cite{Campbell:2011cu}, and which is measured by the CMS and ATLAS collaborations \cite{ATLAS:2013wla,TheATLAScollaboration:2013zha,CMS:eya} is the transverse mass variable, $M_T$ defined in \Eq{eq:mt}.

\begin{figure}
\begin{center}
\includegraphics[width=6.8cm]{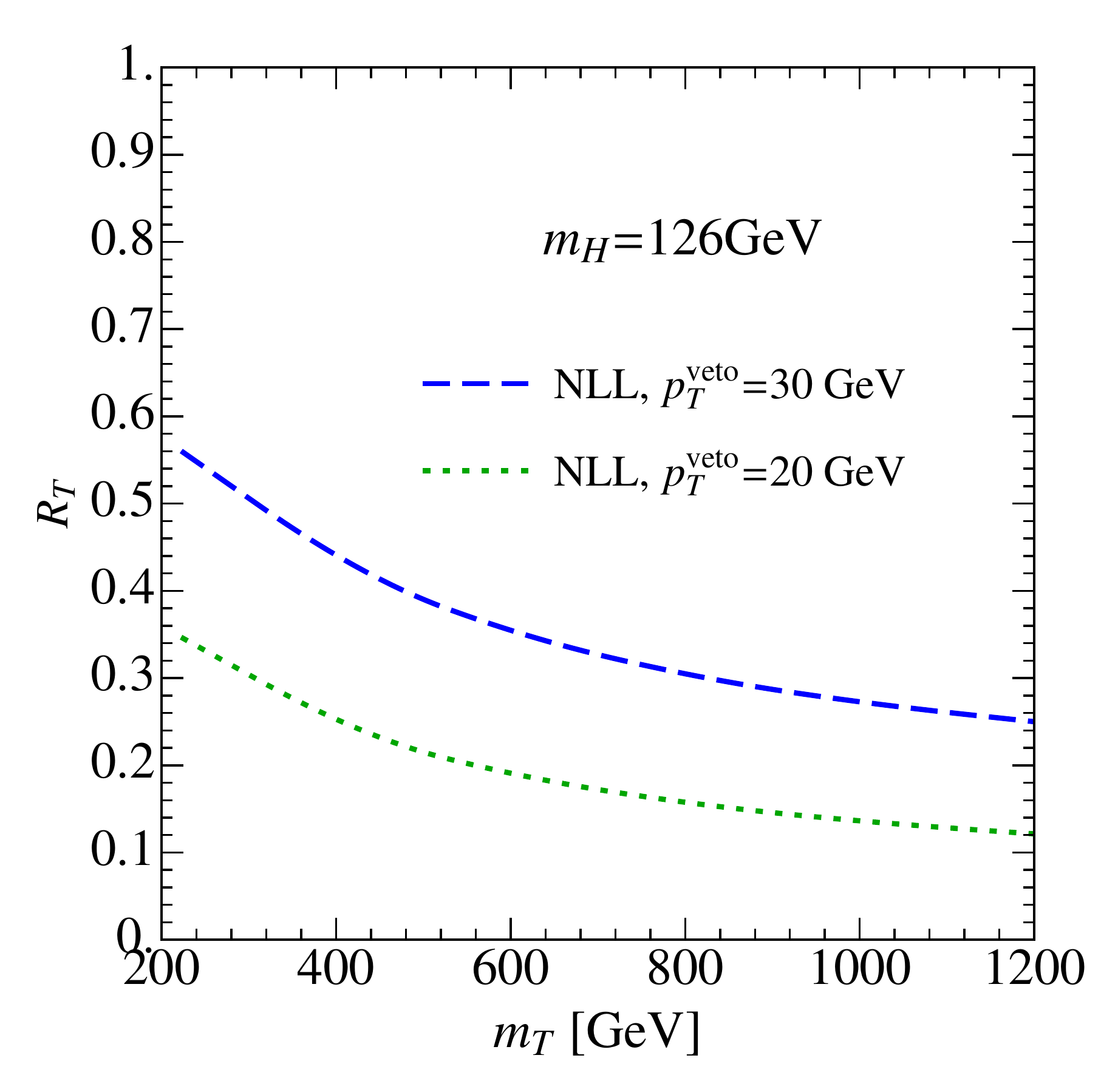}
\end{center}
\caption{Suppression of the exclusive zero jet cross section for off-shell Higgs production as a function of $M_T$. The Sudakov factor is controlled by $\sqrt{\hat s}$, but since $\sqrt{\hat s} \geq M_T$, a larger suppression is observed as a function of $M_T$.} \label{fig:MT}
\end{figure}

Similarly to the ratios considered in \Sec{general} of the exclusive zero jet cross section to the total cross section, as a function of $\hat s$, in \Fig{fig:MT}, we plot the variable
\be \label{eq:MTratio}
R_T=\left .   \left (  \frac{  d\sigma_0^{NLL}(\ptveto)  } {dM_T}   \right )   \middle /      \left (  \frac{d\sigma}{dM_T} \right)      \right . ,
\ee
for $gg\rightarrow H\rightarrow WW\rightarrow \bar \nu_\mu \mu e^+ \nu_e$ in the far off-shell region. Since $M_T$ is designed as a proxy for $\hat s$, the behaviour is as expected from the discussion of \Sec{general}, however, since $\sqrt{\hat s}\geq M_T$, the events contributing at a given $M_T$ all have a larger $\sqrt{\hat s}$. Since it is the $\sqrt{\hat s}$ that governs the Sudakov suppression due to the jet veto, the suppression due to the jet veto at a given $M_T$ is larger than at the same value of $\sqrt{\hat s}$.

We should note that while the values of $M_T$ at which the suppression due to the jet veto becomes significant are larger than is normally considered, or studied experimentally, the authors of \cite{Campbell:2013wga} show that with an improved understanding of the backgrounds in the ATLAS $N_{jet}=0$ bin of the $WW$ data, the $M_T>300$ GeV region, where the jet veto effects are indeed significant, can be used to place a competitive bound on the Higgs width. As will be discussed in \Sec{sec:bound}, their method relies heavily on having an accurate description of the shape of the $M_T$ distribution, which is modified by the jet veto. This section demonstrates that in the exclusive zero jet bin, there is a suppression by a factor of $\sim 2$ above $M_T>300$ GeV, which is a significant effect. This will cause a corresponding weakening of the bound on the Higgs width by a similar factor, which we discuss further in \Sec{sec:bound}.

\subsection{From $8$ TeV to $13$ TeV}\label{sec:higherEcm}
Since the focus will soon shift to the $13$ TeV LHC, in this section we briefly comment on how the effects discussed in the previous sections will be modified at higher $E_{\text{cm}}$. In \Sec{sec:Ecm} we noted that at higher $E_{\text{cm}}$ the jet veto suppression has an increased dependence on $\hat s$ due to the larger range of Bjorken $x$ that is probed in the PDFs. The larger range of available $x$ increases the gluon luminosity at high $\hat s$ allowing for an increased contribution to the cross section from far off-shell effects \cite{Campbell:2013una,Campbell:2013wga}, and increasing the range over which they contribute, potentially amplifying the effects of the jet veto discussed in the previous sections.

\begin{figure}
\begin{center}
\includegraphics[width=8cm]{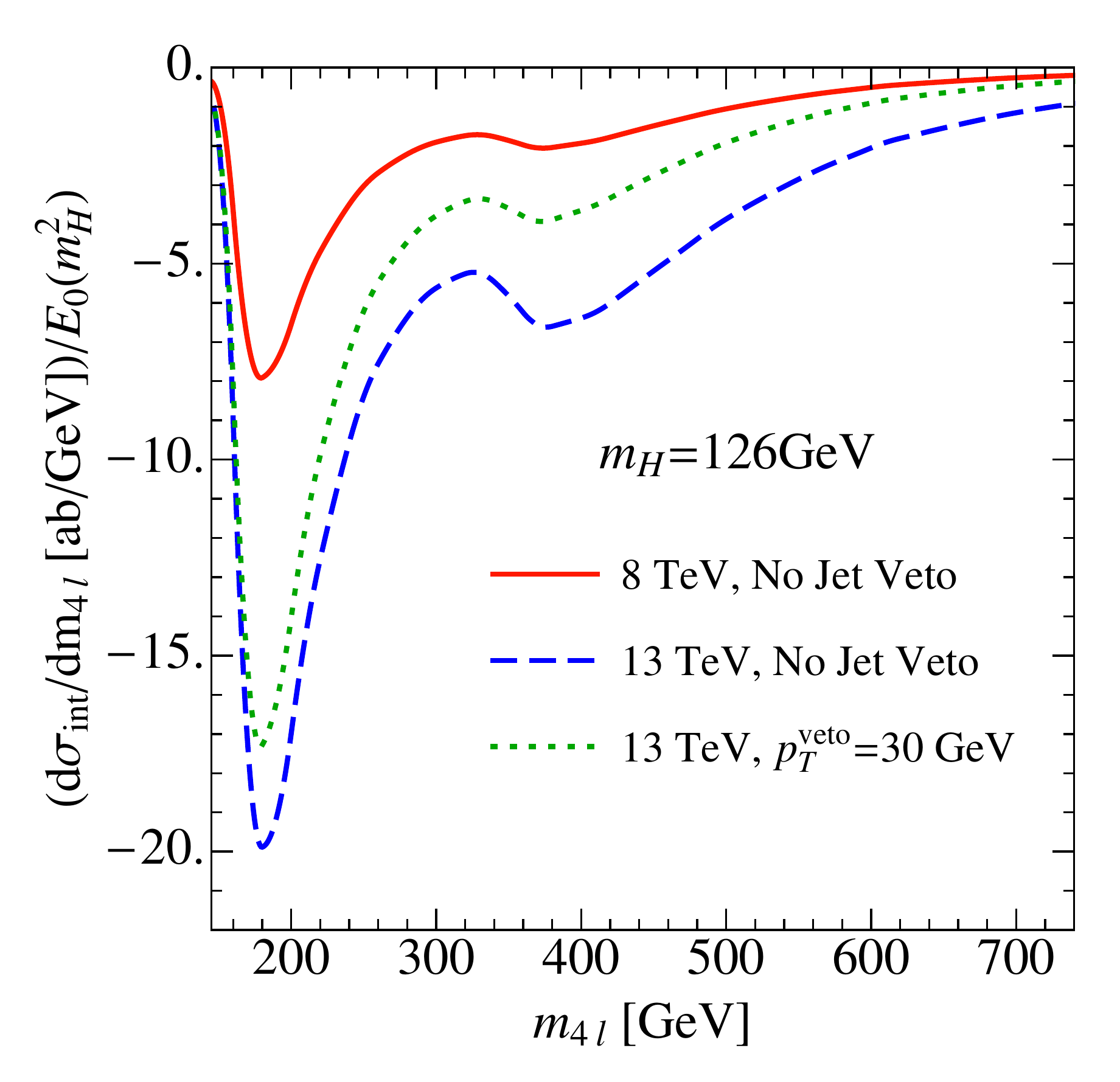}
\end{center}
\caption{A comparison of the signal-background interference, and impact of the jet veto, at $8$, $13$ TeV. At a higher centre of mass energy there is a larger contribution to the cross section from higher $m_{4l}$, where the effect of the exclusive jet veto is largest.}
\label{fig:int_13TeV}
\end{figure}
In this section we consider one example to demonstrate this point, the signal-background interference for $m_H=126$ GeV. The signal-background interference distribution for $E_\text{cm}=8$ and $13\,$TeV is shown in \Fig{fig:int_13TeV}, along with the signal-background interference in the exclusive zero-jet bin at $13$ TeV. As was done in \Fig{fig:int} for the NLL predictions, we have normalized the distribution by the jet veto suppression at $m_H$ so that the suppression is relative to the on-shell production. The most obvious modification compared with $E_\text{cm}=8$ TeV, is the significant enhancement of the signal-background interference cross section, due to the large enhancement of the gluon luminosity at larger $\hat s$. In particular, the contribution to the signal-background interference cross section from the peak at $m_{4l}=2m_t$ is enhanced at higher $E_\text{cm}$, relative to the contribution from $m_{4l}\sim 2m_W$. Since there is a larger relative contribution from higher invariant masses, where the suppression due to the jet veto is larger, the effect of an exclusive jet veto is larger at $13$ TeV. This is in addition to the fact that at $13$ TeV, the $\hat s$ dependence of the suppression due to the jet veto is slightly stronger, as was demonstrated in \Sec{sec:Ecm}.

We again emphasize that when cuts are applied to gain sensitivity to the off-shell region, the effect of the jet veto is not small. In particular, for $13$TeV, there is a significant region above $m_{4l}\sim350$GeV where the suppression due to the jet veto is $\gtrsim 2$, as is seen in \Fig{fig:int_13TeV}.

Although we have focused on the effect of an increased centre of mass energy on a particular observable, the conclusions apply generically, for example, for the $H\rightarrow ZZ\rightarrow 4l, 2l2\nu$ channel, which exhibits similar signal-background interference. Indeed as the centre of mass energy is increased one has the ability to probe phenomena over an increasingly large range of $\hat s$. This amplifies the effects of off-shell physics, as well as the effect of an exclusive jet veto. These effects will be important in any physics channel for which a jet veto is applied, and for which one is interested in the physics over a range of $\hat s$.

\section{Effect of Jet Vetoes on Higgs Width Bounds}\label{sec:bound}

In this section we discuss the impact of jet binning and jet vetoes for the recent program to use the off-shell cross section in $H\rightarrow WW,~ZZ$ to bound the Higgs width \cite{Caola:2013yja,Campbell:2013una,CMS:2014ala,Campbell:2013wga}. Although we have focussed on the case of $H\rightarrow WW$, we will review also the strategy for $H\rightarrow ZZ$ which is similar, also exhibiting a large contribution from the far off-shell cross section, as well as signal-background interference analogous to that in $H\rightarrow WW$. We first discuss the procedure used to bound the Higgs width, and then relate it to our discussion of the suppression of the off-shell cross section in the exclusive zero jet bin. Our focus will be on the  effect of the jet vetoes, rather than carrying out a complete numerical analysis. In particular the proper incorporation of backgrounds, and additional experimental cuts is beyond the scope of this paper.

The method used to bound the Higgs width in Refs.~\cite{Caola:2013yja,Campbell:2013una,Campbell:2013wga} can be phrased in a common language for both $H\rightarrow ZZ, ~WW$. It is based on the different scalings of the on-shell, off-shell and interference contributions to the Higgs cross section, as discussed in \Sec{sec:intro}. Recalling the scaling from \Eq{Eq:Scaling}, the total cross section can be written as
\begin{align}\label{eq:scaling_bound}
\sigma_{H+I}=A+B \left ( \frac{\Gamma_H}{\Gamma_H^{SM}} \right)+C\sqrt{\frac{\Gamma_H}{\Gamma_H^{SM}}},
\end{align}
where the coefficients $A,B,C$ correspond to on-shell, off-shell Higgs mediated, and signal-background interference contributions, respectively. The coefficients depend strongly on the set of cuts that are applied.
To extract a bound on the Higgs width, the procedure is as follows. First, one determines a normalization factor between the experimental data and theoretical prediction, which is as independent as possible of the Higgs width. This can be done for $WW$ by using a strict $M_T$ cut, for example $0.75m_H\leq M_T \leq m_H$, and for $ZZ$ by using a strict cut on $\hat s$. In both cases, this essentially eliminates the coefficients $B,C$ corresponding to the off-shell production and interference. Once this normalization factor is determined, it is scaled out from the entire differential distribution, so that we now must consider the ratio of offshell and onshell production cross sections.  One can then compute the predicted number of events above some $M_T$, or $\hat s$ value, for example, $M_T,~\sqrt{\hat s} \geq 300$ GeV. In this region, the interference and off-shell production dominate the cross section, so that the coefficients $B,C$ are significant, and the expected number of events is sensitive to the Higgs width, as can be seen from the scalings in \Eq{eq:scaling_bound}. By comparing with the number of events observed by the experiment in this region, one can place a bound on the Higgs width. 

This method relies on the ability to normalize the theoretical prediction to data in the low $M_T$, or $\hat s$ region, which is insensitive to the dependence on the Higgs width, and then use the same normalization in the high $M_T$, or $\hat s$ region where there is a large sensitivity to the Higgs width through off-shell production and signal-background interference. However, to be able to do this, one needs to have an accurate theory prediction for the shape of the $M_T$, or $ \hat s $ distribution, particularly in the high $M_T$, or $\hat s$ region. 

As we have seen, the jet veto significantly modifies the shape of the $M_T$, or $\hat s$ distribution, causing it to fall off more rapidly at high $M_T$, or $\hat s$.   Often we presented our results by normalizing the offshell cross sections to the cross section at the Higgs mass. Given the agreement between theory and experiment at $m_H$, this normalization corresponds exactly to what is done if the theory prediction is normalized to the experimental data in the on-shell region, and therefore shows the extent to which a prediction without the inclusion of a jet veto overestimates the contribution to the cross section at high $M_T$  or $ \hat s$ compared with that in the exclusive zero jet bin. 

In both the $H\rightarrow ZZ$, and $H\rightarrow WW$ analyses, jet vetoes or jet binning are used, so it is interesting to consider how they will effect the width bounds. Their use in the two channels is quite different so we will discuss them separately. 

For the case of $H\rightarrow WW$ the jet veto plays an important role because the exclusive zero jet bin dominates the sensitivity, so the jet veto has a large impact on the potential Higgs width bound. This is because effectively it is more difficult to recover the jets which migrate out of the zero jet bin. The plots of the off-shell distributions in \Sec{sec:WW} show the extent to which a prediction without the inclusion of a jet veto overestimates the contribution to the cross section at high $M_T$. This will lead to a weakening on the bound of the Higgs mass, compared with a calculation that does not incorporate the effect of the jet veto. For example, in \cite{Campbell:2013wga}, which first proposed the use of the $H\rightarrow WW$ channel, the estimated sensitivity was derived by comparing an inclusive calculation for the off-shell cross section with data in the exclusive zero jet bin. Here the effect of the restriction to the zero jet bin is not small, and will worsen the bounds by a factor of $\sim 2$, as can be seen by the suppression of the far off-shell cross section in the exclusive zero jet bin, shown in \Sec{sec:WW}.  In an analogous experimental analysis this Sudakov suppression from the jet veto will be accounted for up to some level of precision by the use of a parton shower.  Because this is such a large effect, we believe that an experimental analysis of the high $M_T$ region performed to bound the Higgs width, would benefit from using an analytic calculation of the jet veto suppression in the exclusive zero jet bin, instead of relying on the parton shower. We have demonstrated that the resummation, including the signal-background interference, can be achieved to NLL. Once the NLO virtual corrections are calculated for the interference, these results can also easily be extended to NNLL.

In the $H\rightarrow ZZ\rightarrow 2l2\nu$ channel the situation is different, as the jet binning procedure is used to optimize sensitivity, splitting the data into exclusive 0-jet and inclusive 1-jet categories with comparable bounds coming from each category~\cite{CMS:2014ala}.  Because the inclusive 1-jet channel is still experimentally clean, the large migration to the inclusive 1-jet bin discussed in \Sec{sec:1jet} should have a small (or no) impact on the width bounds derived from the combination of the two channels in $H\to ZZ$. A proper treatment of the migration of events with changing $\hat s$ is still important when considering any improvement that is obtained by utilizing jet binning.  The analytic results for the Sudakov form factor discussed here for $H\to WW$ could be utilized for jet bins for $H\to ZZ$ in a straightforward manner.

\section{Conclusions}\label{sec:conclusion}

In this paper a study of the effect of jet vetoes on off-shell cross sections was made. A factorization theorem in SCET allowed us to analytically treat the summation of Sudakov logarithms, and make a number of general statements about the effect of the jet veto. In particular, the restriction on radiation imposed by the jet veto causes a suppression to the exclusive 0-jet cross section, and correspondingly an enhancement of the inclusive 1-jet cross section, which depends strongly on $\hat s$. For gluon initiated processes the $\hat s$ dependence of the suppression is greater than for quark initiated processes, which is important for channels where the signal is dominated by one production channel, and the background by another. 

The fact that the jet veto suppression is $\hat s$ dependent has interesting effects on differential distributions in $\hat s$, as well as on signal-background interference. To demonstrate these effects, we considered the $gg\rightarrow H\rightarrow WW$ process, which has large off-shell contributions as well as signal-background interference. We performed an NLL resummation for the $gg\rightarrow H\rightarrow WW\rightarrow l\nu l\nu$ process, including a discussion of the resummation for the signal-background interference, for $m_H=126$ GeV, and $m_H=600$ GeV. These two examples demonstrated that depending on the structure of the interference, the jet veto can either enhance or suppress interference effects relative to the on-shell production. For a low mass Higgs a suppression is observed, while for a high mass Higgs there is a significant enhancement in the interference. These effects must be properly incorporated in high mass Higgs searches that use jet vetoes.

The modification of differential distributions in $\hat s$ or $M_T$ due to the $\hat s$ dependence of the jet veto suppression is particularly relevant to a recent program to bound the Higgs width using the off-shell cross section \cite{Caola:2013yja,Campbell:2013una,Campbell:2013wga,CMS:2014ala}. In particular, for the $H\rightarrow WW$ channel, where an exclusive 0-jet veto is imposed to mitigate large backgrounds, the jet veto weakens the bound on the Higgs width by a factor of $\sim 2$ compared to estimates without accounting for the jet veto. Furthermore, since the suppression in the exclusive 0-jet bin corresponds to an enhancement in the inclusive 1-jet bin, and the migration is significant as a function of $\sqrt{\hat s}$ a proper understanding of the effect of the jet veto is crucial for experimental analyses which use jet binning.  
This migration may for example play some role in $H\rightarrow ZZ\rightarrow 2l2\nu$, which was recently used by CMS to place a bound on the off-shell Higgs width, and which uses jet binning in exclusive 0-jet and inclusive 1-jet bins \cite{CMS:2014ala}. 

We presented a factorization theorem in SCET which allows for the resummation of large logarithms of $\sqrt{\hat s}/ \ptveto$, including for the signal-background interference, in a systematically improveable manner. This allows for the analytic study of the effect of the jet veto on the exclusive 0-jet and inclusive 1-jet bins, including the correlations in their theory uncertainties. A complete NNLL calculation would require the calculation of the NLO virtual corrections to the interference, but would allow for the analytic incorporation of jet radius effects, and would place the study of the off-shell cross section on a firmer theoretical footing. Furthermore, since our hard functions are fully differential in leptonic momenta, realistic experimental cuts on the leptonic phase space can be easily implemented. We leave a more detailed investigation, including the treatment of such cuts, and a calculation of the effect of the jet veto on the backgrounds, for future study.

With the LHC beginning its $13$ TeV run in the near future, the importance of the effects discussed in this paper will be amplified. As the centre of mass energy is raised, the range of $\hat s$ which can be probed increases. This typically increases the importance of off-shell effects, as well as the impact of the jet veto, which is essential for an accurate description of the differential distribution in $\hat s$.  In general a proper theoretical understanding of jet vetoes and jet binning for large $\hat s$ can be achieved through resummation, and is important when theoretical cross sections are needed for the interpretation of experimental results.

\begin{acknowledgments}
We thank Frank Tackmann, Jesse Thaler, Markus Klute, and Andrew Larkoski for useful discussions.  We thank the Erwin Schroedinger Institute and the organizers of the ``Jets and Quantum Fields for LHC and Future Colliders'' workshop for hospitality and support while portions of this work were completed. This work was supported in part by the Office of Nuclear Physics of the U.S. Department of Energy under the Grant No. DE-FG02-94ER40818. I.M. is also supported by NSERC of Canada. 
\end{acknowledgments}

\bibliography{WWbib}{}
\bibliographystyle{jhep}

\end{document}